%% file: boundsUpdatePaper.tex
\documentclass[preprint,prd,11pt]{revtex4}
\usepackage{amsmath,amssymb,graphicx}
\usepackage{xspace}
\usepackage{setspace}
\usepackage{ulem,color}
\usepackage[mathscr]{eucal}
\usepackage{array}
\usepackage{slashed}
\usepackage{float}
\usepackage{url}
\usepackage{textcomp}
\usepackage{lscape}
\usepackage{wrapfig}
\usepackage{feynmp}
\usepackage{hyperref}
\newcommand{\SARAH}{{\tt SARAH}\xspace}
\newcommand{\Bs}{B^0_s}
\newcommand{\Bd}{B^0_d}
\newcommand{\SPheno}{{\tt SPheno}\xspace}

\newcommand{\abs}[1]{\left|{#1}\right|}
\newcommand{\eps}{\epsilon}
\newcommand{\ampM}[0]{\mathcal{M}}

\newcommand{\kl}[1]{\left(#1\right)}
\newcommand{\Ekl}[1]{\left[#1\right]}

\newcommand{\pprime}{{\prime\prime}}
\newcommand{\nn}{\nonumber}

\newcommand{\scn}[2]{\ensuremath{#1\cdot 10^{#2}}}

 \newcommand{\Bsmm}{\ensuremath{B^0_s\to \mu\bar{\mu}\xspace}}
 
\newcommand{\Bsdll}{\ensuremath{B^0_{s,d}\to \ell \bar{\ell} \xspace}}

\newcommand{\Bdmm}{\ensuremath{B^0_d\to \mu\bar{\mu}\,}}
\newcommand{\Bsdmm}{\ensuremath{B^0_{s,d}\to \mu\bar{\mu}\,}}

\newcommand{\bsgamma}{\ensuremath{B\to X_s \gamma\,}}

\newcommand{\BRx}[1]{\ensuremath{\text{BR}(#1)\,}}
\newcommand{\BR}{\ensuremath{\text{BR}\,}}

\newcommand{\sfd}[1]{{\bf #1}}
\newcommand{\sfb}[1]{{\bf {\bar #1}}}

 \newcommand{\RpV}[0]{\ensuremath{R\mathrm{pV}}\xspace}
\newcommand{\BnV}{\slashed B}
\newcommand{\LnV}{\slashed L}

\newcommand{\DRbar}{\ensuremath{\overline{\mathrm{DR}}}}

\newcommand{\tanb}{\ensuremath{\tan\beta}}
\newcommand{\GeV}{\text{Ge}\hspace{-0.05cm}\text{V}}
\newcommand{\TeV}{\text{Te}\hspace{-0.05cm}\text{V}}
\newcommand{\lpa}[2]{\ensuremath{\lambda^{\prime *}_{#1}\lambda^\prime_{#2}}}
\newcommand{\lpp}[2]{\ensuremath{\lambda^{\prime\prime *}_{#1}\lambda^{\prime\prime}_{#2}}}
\newcommand{\spa}[1]{\ensuremath{\widetilde{#1}}}
\newcommand{\enquote}[1]{``#1''}

\newcommand{\AddrBonn}{%
Bethe Center for Theoretical Physics \& Physikalisches Institut der 
Universit\"at Bonn, \\
Nu{\ss}allee 12, 53115 Bonn, Germany
}

\graphicspath{ 
{./plots/}
}

\preprint{BONN-TH-2013-14}
\begin{document}

\title{$B_{s,d}^0 \to \mu\bar{\mu}$ and $B\to X_s\gamma$ in the $R$-parity violating MSSM}

\author{H. K. Dreiner}\email{\tt dreiner@uni-bonn.de}
\author{K.~Nickel}\email{\tt nickel@th.physik.uni-bonn.de}
\author{F. Staub}\email{\tt fnstaub@physik.uni-bonn.de}
\affiliation{\AddrBonn}


\begin{abstract}

  The recent measurements of $\Bsmm$ decay candidates at the LHC
  consistent with the standard model rate, and the improving upper
  limits for $\Bdmm$ can strongly constrain beyond the standard model
  physics. For example, in supersymmetric models with broken
  $R$-parity (RpV), they restrict the size of the new couplings. We use the
  combination of the public software packages \SARAH and \SPheno to
  derive new bounds on several combinations of $\RpV$ couplings.  We
  improve existing limits for the couplings which open tree-level
  decay channels and state new limits for combinations which induce
  loop contributions. This is the first study which performs a full
  one-loop analysis of these observables in the context of $R$-parity
  violation. It turns out that at one-loop despite the strong
  experimental limits only combinations of $R$-parity violating
  couplings are constrained which include third generation
  fermions. We compare our limits with those obtained via $B\to
  X_s\gamma$ and discuss the differences.
\end{abstract}


\maketitle

\input{introduction}
\input{Bdecays}
\input{model}

\input{setup}

\input{results}

\input{conclusion}

\section*{Acknowledgements}
We thank Werner Porod and Manuel Krauss for helpful discussions.


\input{lit.tex}

\bibliographystyle{utphys}

\end{document}

%% file: introduction.tex
\section{Introduction}
The first experimentation phase of the experiments at the Large Hadron Collider (LHC) is 
completed. However, there is no evidence or hint up to now for superpartner particles as 
predicted by the well-motivated theory of supersymmetry (SUSY) or any other physics 
beyond the Standard Model (SM) \cite{:2012rz,Chatrchyan:2012jx,CMS-PAS-SUS-11-022,CMS-PAS-SUS-12-005,:2012mfa}.  
The simplest SUSY scenarios like the constrained minimal supersymmetric standard 
model (CMSSM) is under pressure by the ongoing non-discovery, leading to the 
exclusion of large areas of parameter space \cite{Bechtle:2012zk,Ghosh:2012dh,Buchmueller:2012hv}.  
In addition, the observed mass of $m_h\approx 126~{\GeV}$ for the Higgs boson 
\cite{Atlas:2012gk,CMS:2012gu} is rather hard to realize in the CMSSM and requires 
heavy SUSY spectra to push the predicted Higgs mass to that level \cite{pMSSM,Bechtle:2012zk,Buchmueller:2012hv}.  
However, this applies only if the stop and the other sfermion masses are related. While 
heavy stops with a large mass splitting are needed to explain the Higgs mass, the other 
sfermion contributions are usually sub-dominant in this context. Hence, these states 
could in principle be much lighter. However they are constrained by direct searches. 
Therefore SUSY models with different signatures like $R$-parity violation (RpV) are more 
interesting since they can significantly soften the mass limits \cite{Hall:1983id,rpv,Barbier:2004ez,rpvsearches,Franceschini:2012za,Evans:2012bf} 
and provide a rich collider phenomenology \cite{Dreiner:2012wm}.  On the other hand, 
beyond the standard model (BSM) physics can not only manifest itself directly at collider 
searches, but also indirectly via quantum corrections to (rare) standard model processes. 
Interesting processes are those which rarely occur in the SM but which can be measured 
with high accuracy. In this context quark flavor changing neutral currents (qFCNC), like 
$B\to X_s\gamma$ \cite{Barbieri:1993av,bsgamma1,bsgamma2,BsGamma} and the decays of the neutral $B^0$ mesons ($B_s^0,B_d^0$) 
into a pair of leptons \cite{bsmumu-susy} are interesting candidates to look for deviations 
from the SM.

In this paper we focus on the constraints on $R$-parity violating couplings derived from the experimental limits on $B^0_{s,d}\to \mu \bar \mu$ and $B\to X_s\gamma$. Previous studies of $B\to X_s\gamma$ in this context assumed a SUSY spectra no longer in agreement with experimental data \cite{deCarlos:1996yh,Kong:2004cp}, while for $B$-meson decays to two leptons only the new tree level contributions have been studied so far \cite{Dreiner:2006gu,Li:2013fa,Yeghiyan:2013upa}. We perform a full one-loop analysis of $B^0_{s,d}\to \mu \bar \mu$ which allows us to constrain new combinations of couplings besides those at tree level. For this purpose we use the combination of public software packages \SARAH \cite{sarah} and \SPheno \cite{spheno}. \SARAH creates new source code for \SPheno which can be used for the numerical study of a given model.  Recently, this functionality has been extended to provide a full one-loop calculation of $B^0_{s,d}\to \ell \bar \ell$ \cite{Dreiner:2012dh}. We compare the new limits with those obtained by a revised study of $B\to X_s\gamma$ and discuss the differences between both observables.

We briefly review the main basics of the $B$-decays in sec.~\ref{sec:Bdecays} and introduce the MSSM with $R$-parity violation in sec.~\ref{sec:MSSMRpV}. We explain the numerical setup in sec.~\ref{sec:setup} and present our results in sec.~\ref{sec:results}, before concluding in sec.~\ref{sec:conclusion}.

%% file: Bdecays.tex
\section{Standard model predictions and measurements for $B_{s,d}^0 \to \mu\bar{\mu}$ and $B\to X_s\gamma$}
\label{sec:Bdecays}
\subsection{$B_{s,d}^0 \to \mu\bar{\mu}$}
The semi-leptonic $B^0$ decay is described by a matrix element $\mathcal{M}$ as a function of the form factors $F_S,F_P,F_V,F_A$ for the scalar, pseudoscalar, vector and axial vector currents. The squared matrix element \cite{Dedes:2008iw} of $B_q^0\to \ell_k\bar{\ell}_l$,
\begin{align}
  \label{eq:squaredMBsllp}
  (4\pi)^4 \abs{\ampM}^2&=2\abs{F_S}^2\Ekl{M_{B^0_q}^2-(m_l+m_k)^2}
+2\abs{F_P}^2\Ekl{M_{B^0_q}^2-(m_l-m_k)^2}
  \\
\nn &+ 2\abs{F_V}^2\Ekl{M_{B^0_q}^2(m_k-m_l)^2-(m_k^2-m_l^2)^2} \\
\nn &+ 2\abs{F_A}^2\Ekl{M_{B^0_q}^2(m_k+m_l)^2-(m_k^2-m_l^2)^2} \\
\nn &+ 4\Re (F_s F_V^*) (m_l-m_k)\Ekl{M_{B^0_q}^2+(m_k+m_l)^2} \\
\nn &+ 4\Re (F_P F_A^*) (m_l+m_k)\Ekl{M_{B^0_q}^2-(m_k-m_l)^2} \, ,
\end{align}
determines the branching ratio $\BR(B_q^0\to \ell_k\bar{\ell}_l)$ \cite{Dedes:2008iw},
\begin{equation}
  \label{eq:Bsllpbranching}
  \BR(B_q^0\to \ell_k\bar{\ell}_l)=\frac{\tau_{B^0_q}}{16\pi}
  \frac{\abs{\mathcal{M}}^2}{M_{B_q^0}}\sqrt{1-\kl{\frac{m_k+m_l}{M_{B_q^0}}}^2}\sqrt{1-\kl{\frac{m_k-m_l}{M_{B_q^0}}}^2},
\end{equation}
where $\tau_{B^0_q}$ is the lifetime of the mesons and $m_k$ the mass of the lepton $\ell_k$. Note that the form factor $F_V$ does not contribute to Eq.~(\ref{eq:squaredMBsllp}) in the case $l=k$.
These decays are fixed in the SM by the CKM matrix and the form factors can be calculated with a high precision. The predicted branching ratios for $B^0_{s,d} \to \mu \bar{\mu}$ are \cite{Buras:2012ru}. 
\begin{eqnarray}
  \BRx{B^0_s \to \mu \bar{\mu}}_{\text{SM}} &=& \scn{(3.23\pm 0.27)}{-9} \label{eq:Bsintro1}, \\
  \BRx{B^0_d \to \mu \bar{\mu}}_{\text{SM}} &=& \scn{(1.07\pm 0.10)}{-10}.
\end{eqnarray}
The errors include experimental uncertainties of the involved parameters as well as uncertainties 
from higher orders and scheme dependence.
These predictions neglect the CP violation in the $B_s$--$\bar B_s$
system which leads to a difference in the decay widths for $B^0_s$ and
$\bar{B}^0_s$ \cite{Raven:2012fb}. When it is not known in the
experiment whether a pair of muons comes from the decay of a $B_s^0$
or a $\bar B_s^0$, the untagged decay rate is measured. Therefore, one
has to compare the LHC limits with the averaged branching ratio of
$B^0_s \to \mu \bar{\mu}$ and $\bar B^0_s \to \mu \bar{\mu}$
\cite{Buras:2013uqa}
\begin{align}
  \overline{\BR}({B_s^{0}\to\mu\bar \mu})_{\text{SM}} &= \scn{(3.56\pm 0.18)}{-9} \label{eq:Bsuntagged} 
\end{align}
The width difference between $B_d^0$ and $\bar B_d^0$ is much smaller
than for the $B_s$ system and is not measured
accurately. Hence, we use the untagged
rate in the following. Eq.~(\ref{eq:Bsuntagged}) is consistent with
the recently updated measurements for $B_s\to \mu\bar \mu$ at the LHC
\cite{LHCb:2012ct,LHCb:2013},
\begin{align}
  \label{eq:LHCbbounds}
  \BRx{B^0_s \to \mu \bar{\mu}} &= (2.9^{+1.1}_{-1.0})\times 10^{-9}, 
\end{align}
In addition, the experimental upper limit for \cite{LHCb:2013}
\begin{align}
    \label{eq:LHCbbounds2}
  \BRx{B^0_d \to \mu \bar{\mu}} &< 7.4\times 10^{-10},
\end{align}
is approaching the SM
expectation. These measurements shrink the space where one can hope to
see new physics. Especially SUSY scenarios with large $\tan\beta$ can
lead to a prediction of these decays which is now ruled out \cite{Haisch:2012re}. 

To compare the bounds of these two observables with our calculation and to put limits on the SUSY contributions, we consider the ratio
\begin{align}
R_i & \equiv \frac{\BRx{B_i^0\to \mu \bar{\mu}}_{\text{SUSY}}}{\BRx{B_i^0\to \mu \bar{\mu}}_{\text{SM}}},\quad (i=s,d),
\end{align}
in which the finite width effects factor out. Note, $\text{BR}_{\text{SUSY}}$ includes also 
the SM contributions. Together with Eqs.~(\ref{eq:LHCbbounds}) and (\ref{eq:LHCbbounds2}), we obtain an allowed range of
\begin{eqnarray}
\label{eq:ranges}
& 0.43 < R_s < 1.35 \, , &\\
& R_d < 8.30 \,  . &
\end{eqnarray}
Here, we assumed a combined total uncertainty of 20\% on the upper (and lower) limit, which includes the errors of the SM prediction and of our  SUSY calculation. 
\subsection{$B\to X_s\gamma$}
The main contribution to the radiative B-meson decay
$\bar{B} \rightarrow X_s \gamma$ stems from the partonic process \(b
\rightarrow s \gamma\). The standard model prediction \cite{bsgamma1,bsgamma2,BsGamma}
\begin{equation}
\text{Br}(B \to X_s \gamma)_{SM} = (3.15 \pm 0.23) \times 10^{-4},
\end{equation}
has to be compared with the experimental limit of \cite{Beringer:1900zz}
\begin{equation}
\text{Br}(B \to X_s \gamma) = (3.55 \pm 0.24 \pm 0.09) \times 10^{-4}  \, .
\end{equation}
To derive bounds on the $\RpV$ couplings from $B \to X_s\gamma$ we follow closely the approach of Ref.~\cite{Haisch:2012re} and use as 95\% C.L. limit
\begin{equation}
0.89 < R_{X_s \gamma} < 1.33 \, ,
\end{equation}
with 
\begin{equation}
R_{X_s \gamma} \equiv  \frac{\BRx{B \to X_s \gamma}_{SUSY}}{\BRx{B \to X_s \gamma}_{SM}}.
\end{equation}

%% file: model.tex
\section{$R$-parity violation and neutral B-meson decays}
\label{sec:MSSMRpV}
$R$-parity is a discrete $Z_2$ symmetry of the MSSM which is defined as
\begin{equation}
  \label{eq:RParity}
  R_P = (-1)^{3(B-L)+2s} \,,
\end{equation}
where $s$ is the spin of the field and $B$, $L$ are its baryon
respectively lepton number.  If we just allow for $R$-parity
conserving parameters and assume the minimal set of superfields which
is anomaly free and needed to reproduce the SM, we are left with the
(renormalizable) superpotential of the MSSM
\begin{equation}
 W_R= Y^{a b}_{e}\,\sfd L_a \sfb E_b \sfd H_d 
   +Y^{a b}_{d}\, \sfd Q_a \sfb D_b \sfd H_d + 
   Y^{a b}_{u} \sfd Q_a \sfb U_b \sfd H_u +
 \mu\, \sfd H_u \sfd H_d \, .
\end{equation}
Here $a,b=1,2,3$ are generation indices, while we suppressed
 color and isospin indices.  The corresponding standard
soft-breaking terms for the scalar fields $\spa L,\spa E,\spa
Q, \spa U,\spa D, H_d, H_u$ and the gauginos
$\spa{B},\spa{W},\spa{g}$ read
\begin{eqnarray}
\nonumber -\mathscr{L}_{\text{SB},R} &=& m_{H_u}^2 |H_u|^2 + m_{H_d}^2
|H_d|^2+ \spa{Q}^\dagger m_{\spa{q}}^2 \spa{Q} +
\spa{L}^\dagger m_{\spa{l}}^2 \spa{L} + \spa{D}^\dagger
m_{\spa{d}}^2 \spa{D} + \spa{U}^\dagger m_{\spa{u}}^2 
\spa{U} + \spa{E}^\dagger m_{\spa{e}}^2 \spa{E} \nonumber \\ && + \frac{1}{2}\left(M_1 \, \spa{B}
\spa{B} + M_2 \, \spa{W}_i \spa{W}^i + M_3 \, \spa{g}_\alpha
\spa{g}^\alpha + h.c.\right) \nonumber \\ && 
+ (\spa{Q}
T_u\spa{U}^\dagger H_u +  \spa{Q} T_d \spa{D}^\dagger H_d + 
\spa{L} T_e \spa{E}^\dagger H_d + B_\mu H_u H_d + \text{h.c.}) \thickspace .
\end{eqnarray}
However, there are additional {renormalizable interactions which are
 allowed by gauge invariance in the superpotential but
break $R$-parity:
\begin{align}
  \label{eq:superpotRpV}
  W_{\text{tri}\slashed L}&=\frac 12 \lambda_{ijk} \cdot \sfd L_i\sfd L_j \sfb E_k+\lambda_{ijk}^\prime\cdot \sfd
    L_i \sfd Q_j \sfb D_k , \\
   \label{eq:superpotRpVbi} 
  W_{\text{bi} \slashed L}&=\kappa_i \sfd L_i \sfd H_u,\\
 \label{eq:superpotRpVbnv} 
 W_{\slashed B}&= \frac 12\lambda_{ijk}^{\prime\prime} \sfb U_i \sfb D_j \sfb D_k. 
\end{align}
The bi- and trilinear operators in eqs.~(\ref{eq:superpotRpV}) and (\ref{eq:superpotRpVbi}) violate  lepton number, the operators in  eq.~(\ref{eq:superpotRpVbnv}) violate baryon number. The corresponding soft-breaking terms for these interactions are 
\begin{align}
  \label{eq:soft}
  -\mathscr{L}_{\text{tri}\slashed L}&=\frac 12 T_{\lambda_{ijk}} \cdot \spa L_i \spa L_j \spa E_k+T_{\lambda_{ijk}^\prime}\cdot \spa
    L_i \spa Q_j \spa D_k + \text{h.c.} \,, \\
  -\mathscr{L}_{\text{bi} \slashed L}&=B_{\kappa_i} \spa L_i H_u + \text{h.c.} \,,\\
 -\mathscr{L}_{\slashed B}&= \frac 12 T_{\lambda_{ijk}}^{\prime\prime} \spa U_i^\alpha \spa D_j^\beta \spa D_k^\gamma \eps_{\alpha\beta\gamma} + \text{h.c.} \,. 
\end{align}
The terms involving $\lambda$ (or $\lambda^\prime,\lambda^{\pprime}$)
are called {\bf LLE} interactions (or {\bf LQD, UDD}) in the
following. Since proton decay is always triggered by a combination of
baryon and lepton number violating couplings, a model
with either $\LnV$ or $\BnV$ terms is safe from rapid proton decay.
We are going to study in the following the impact of the new couplings
present in $W_{\text{tri}\slashed L}$ and $W_{\slashed B}$ on neutral
B-meson decays. The bilinear terms $\kappa_i$ can be absorbed in
$\lambda$ by a redefinition of the superfields at a given
scale \cite{Hall:1983id,Dreiner:2003hw}.
$\lambda$ and $\lambda^\pprime$ are antisymmetric in the first
  two indices,
\begin{align}
  \lambda_{ijk}&=-\lambda_{jik},\qquad \lambda^\pprime_{ijk}=-\lambda^\pprime_{ikj},
\end{align}
leaving nine independent components to $\lambda$ and $\lambda^\pprime$. The $\lambda^\prime$ tensor has 27 independent components. However, only specific combinations of the parameters can significantly enhance the B-meson decay rate, which do not rely on sub-dominant sfermion flavor mixing:
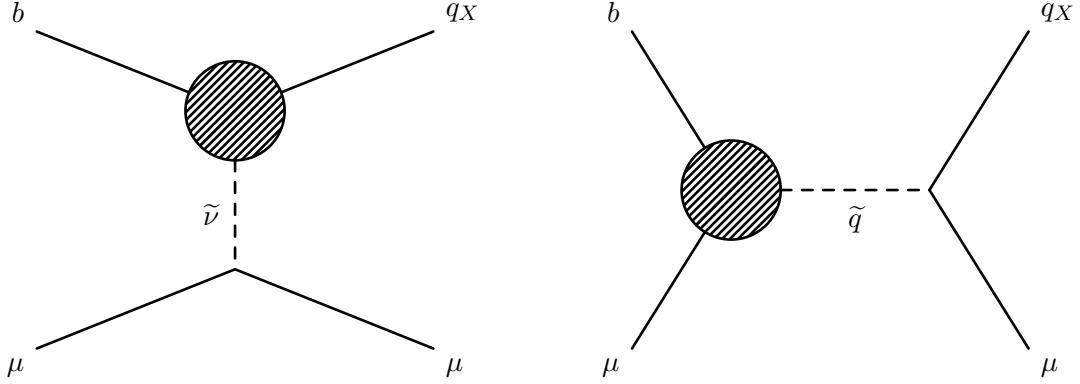
\begin{figure}[bt]
\begin{center}
\begin{fmffile}{diagrams/Tree2}
  \fmfframe(0,0)(0,0){
    \begin{fmfgraph*}(150,150)
    \fmftop{l1,l2}
    \fmfbottom{r1,r2}
    \fmf{plain}{l1,v1,l2}
    \fmf{plain}{r1,v2,r2}
    \fmf{dashes, label=$\spa{\nu}$}{v1,v2}
    \fmfblob{.25w}{v1}
    \fmflabel{$b$}{l1}
    \fmflabel{$q_X$}{l2}
    \fmflabel{$\mu$}{r2}
    \fmflabel{$\mu$}{r1}
\end{fmfgraph*}}
\end{fmffile}
\hspace{2cm}
\begin{fmffile}{diagrams/Tree1}
  \fmfframe(0,0)(0,0){
    \begin{fmfgraph*}(150,150)
    \fmftop{l1,l2}
    \fmfbottom{r1,r2}
    \fmf{plain}{l1,v1}
    \fmf{plain}{l2,v2}
    \fmf{dashes, label=$\spa{q}$}{v1,v2}
    \fmf{plain}{r1,v1}
    \fmf{plain}{v2,r2}
    \fmfblob{.25w}{v1}
    \fmflabel{$b$}{l1}
    \fmflabel{$q_X$}{l2}
    \fmflabel{$\mu$}{r2}
    \fmflabel{$\mu$}{r1}
\end{fmfgraph*}}
\end{fmffile}
\end{center}
\caption{Possible $\RpV$ contributions to $B^0_{s,d}\to \mu\bar{\mu}$
  with SUSY particles in the propagator ($q_X=d,s$ quarks for
  $X=1,2$). These diagrams can cause direct tree level contributions
  $\propto \lambda_{i22}^*\lambda^\prime_{iX3}, \,
  \lambda_{i22}^*\lambda^\prime_{i3X}$ (for the left diagram) or
  $\propto \lambda^{\prime *}_{2iX}\lambda^\prime_{2i3}$ (for the
  right diagram). However, also indirect tree level contributions are possible if one takes the flavor change in the SM or other SUSY-loops into account. The blobs represent all one-loop diagrams.}
\label{fig:DiagramsProp}
\end{figure}

\begin{figure}[bt]
\begin{tabular}{ccccccc}
\begin{fmffile}{diagrams/Zpenguin}
  \fmfframe(10,10)(10,10){
    \begin{fmfgraph*}(100,100)
    \fmftop{l1,l2}
    \fmfbottom{r1,r2}
    \fmf{plain}{l1,v1,l2}
    \fmf{plain}{r1,v2,r2}
    \fmf{wiggly, label=$Z$}{v1,v2}
    \fmfblob{.25w}{v1}
    \fmflabel{$b$}{l1}
    \fmflabel{$\bar{q}_i$}{l2}
    \fmflabel{$\mu$}{r2}
    \fmflabel{$\bar{\mu}$}{r1}
\end{fmfgraph*}}
\end{fmffile}
&
\hspace{1cm}
&
\begin{fmffile}{diagrams/Hpenguin}
  \fmfframe(10,10)(10,10){
    \begin{fmfgraph*}(100,100)
    \fmftop{l1,l2}
    \fmfbottom{r1,r2}
    \fmf{plain}{l1,v1,l2}
    \fmf{plain}{r1,v2,r2}
    \fmf{dashes, label=$\Phi$}{v1,v2}
    \fmfblob{.25w}{v1}
    \fmflabel{$b$}{l1}
    \fmflabel{$\bar{q}_i$}{l2}
    \fmflabel{$\mu$}{r2}
    \fmflabel{$\bar{\mu}$}{r1}
\end{fmfgraph*}}
\end{fmffile}
&
\hspace{1cm}
&
\begin{fmffile}{diagrams/boxes}
  \fmfframe(10,10)(10,10){
    \begin{fmfgraph*}(100,100)
    \fmftop{l1,l2}
    \fmfbottom{r1,r2}
    \fmf{plain}{l1,v1,l2}
    \fmf{plain}{r1,v1,r2}
    \fmfblob{.25w}{v1}
    \fmflabel{$b$}{l1}
    \fmflabel{$\bar{q}_i$}{l2}
    \fmflabel{$\mu$}{r2}
    \fmflabel{$\bar{\mu}$}{r1}
\end{fmfgraph*}}
\end{fmffile}
\end{tabular}
\caption{One-loop contributions to $B^0_{s,d}\to \mu\bar{\mu}$ with SUSY particles only in the loop. The $Z$-penguins (on the left) and Higgs-penguins ($\Phi=h,H,A^0$) are a sum of wave- and vertex-corrections, see also Fig.~\ref{fig:Zpeng}. The diagram on the right represents all possible box contributions. }
\label{fig:diagramsLoop}
\end{figure}
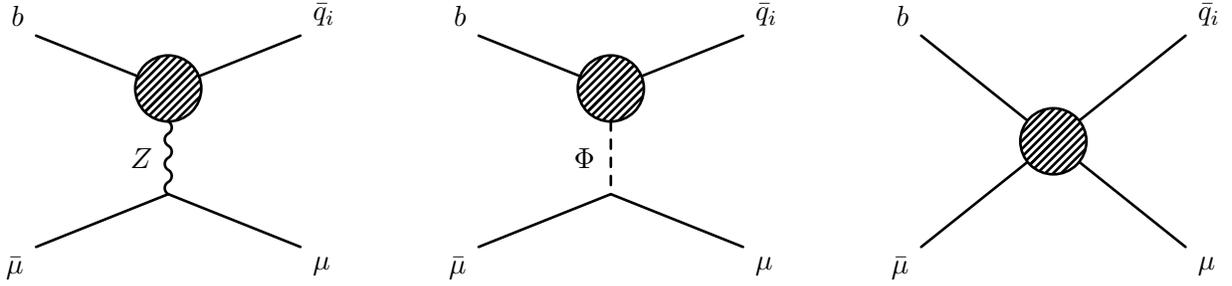

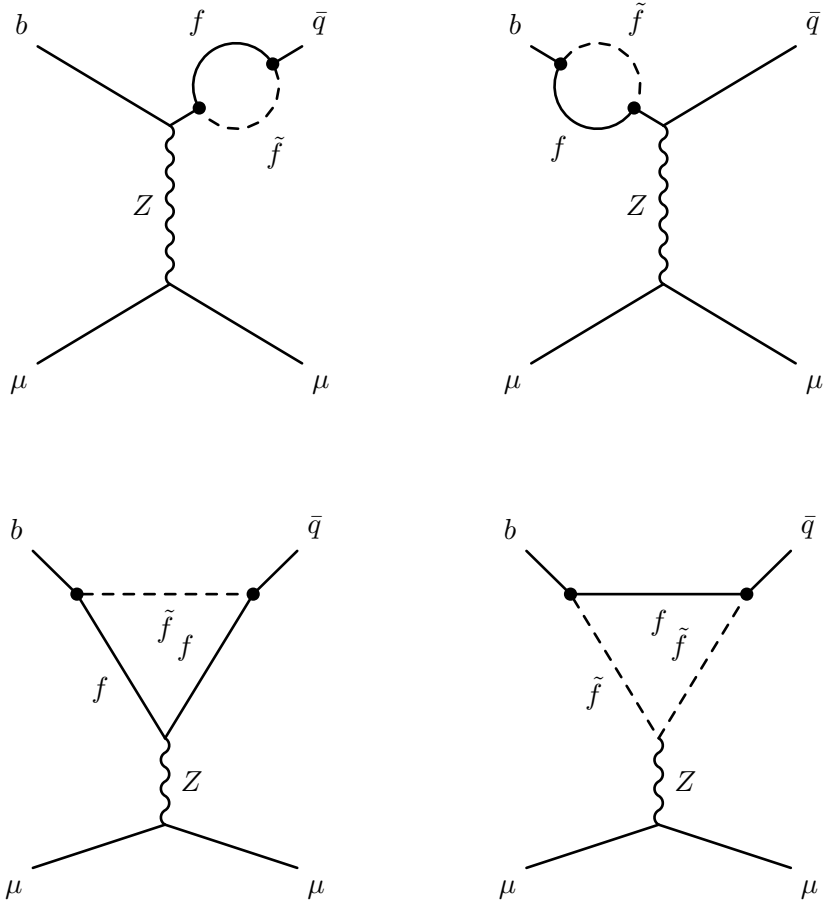
\begin{figure}
\centering
\begin{fmffile}{diagrams/wave1}
  \fmfframe(20,20)(20,20){
\begin{fmfgraph*}(100,150)
\fmftop{l2,l1}
\fmfbottom{r1,r2}
\fmf{plain}{v3,l2}
\fmf{wiggly,label=$Z$}{v3,v4}
\fmf{plain}{v4,r1}
\fmf{plain}{v4,r2}
\fmf{phantom}{l1,v3}
\fmffreeze
\fmf{plain}{l1,v1}
\fmf{plain,right,tension=0.2,label=$f$}{v1,v2}
\fmf{dashes,left,tension=0.2,label=$\tilde{f}$}{v1,v2}
\fmf{plain}{v2,v3}
\fmfdot{v1}
\fmfdot{v2}
\fmflabel{$\bar{q}$}{l1}
\fmflabel{$b$}{l2}
\fmflabel{$\mu$}{r1}
\fmflabel{$\mu$}{r2}
\end{fmfgraph*}}
\end{fmffile}
\hspace{1cm}
\begin{fmffile}{diagrams/wave2}
  \fmfframe(20,20)(20,20){
\begin{fmfgraph*}(100,150)
\fmftop{l1,l2}
\fmfbottom{r1,r2}
\fmf{plain}{v3,l2}
\fmf{wiggly,label=$Z$}{v3,v4}
\fmf{plain}{v4,r1}
\fmf{plain}{v4,r2}
\fmf{phantom}{l1,v3}
\fmffreeze
\fmf{plain}{l1,v1}
\fmf{plain,right,tension=0.2,label=$f$}{v1,v2}
\fmf{dashes,left,tension=0.2,label=$\tilde{f}$}{v1,v2}
\fmf{plain}{v2,v3}
\fmfdot{v1}
\fmfdot{v2}
\fmflabel{$\bar{q}$}{l2}
\fmflabel{$b$}{l1}
\fmflabel{$\mu$}{r1}
\fmflabel{$\mu$}{r2}
\end{fmfgraph*}}
\end{fmffile} \\
\begin{fmffile}{diagrams/penguin1}
  \fmfframe(20,20)(20,20){
\begin{fmfgraph*}(100,150)
\fmftop{l1,l2}
\fmfbottom{r1,r2}
\fmf{plain}{r1,v1}
\fmf{plain}{v1,r2}
\fmf{plain}{l1,v2}
\fmf{plain,label=$f$,tension=0.3}{v2,v3}
\fmf{plain,label=$f$,tension=0.3}{v3,v4}
\fmf{plain}{v4,l2}
\fmf{wiggly,tension=1.0,label=$Z$}{v1,v3}
\fmf{dashes,tension=0.1,label=$\tilde{f}$}{v2,v4} 
\fmfdot{v2}
\fmfdot{v4}
\fmflabel{$\bar{q}$}{l2}
\fmflabel{$b$}{l1}
\fmflabel{$\mu$}{r2}
\fmflabel{$\mu$}{r1}
\end{fmfgraph*}}
\end{fmffile} 
\hspace{1cm}
\begin{fmffile}{diagrams/penguin2}
  \fmfframe(20,20)(20,20){
\begin{fmfgraph*}(100,150)
\fmftop{l1,l2}
\fmfbottom{r1,r2}
\fmf{plain}{r1,v1}
\fmf{plain}{v1,r2}
\fmf{plain}{l1,v2}
\fmf{dashes,label=$\tilde{f}$,tension=0.3}{v2,v3}
\fmf{dashes,label=$\tilde{f}$,tension=0.3}{v3,v4}
\fmf{plain}{v4,l2}
\fmf{wiggly,tension=1.0,label=$Z$}{v1,v3}
\fmf{plain,tension=0.1,label=$f$}{v2,v4} 
\fmfdot{v4}
\fmfdot{v2}
\fmflabel{$\bar{q}$}{l2}
\fmflabel{$b$}{l1}
\fmflabel{$\mu$}{r2}
\fmflabel{$\mu$}{r1}
\end{fmfgraph*}}
\end{fmffile}
\caption{$Z$-penguin diagrams contributing to semi-leptonic $B$-meson decays in case of $R$pV. The particles in the loop are a SM fermion $f$ and a SUSY sfermion $\spa{f}$.  These contributions are proportional either to $\lambda^{\prime *} \lambda^{\prime}$, if leptons/squark or sleptons/quark pairs run in the loop, or to $\lambda^{\pprime *} \lambda^{\pprime}$ if only (s)quarks are involved.}
\label{fig:Zpeng}
\end{figure}

\begin{figure}[bt]
\begin{tabular}{ccc}
\begin{fmffile}{diagrams/bsgamma1}
 \fmfframe(0,0)(0,0){
   \begin{fmfgraph*}(120,120)
    \fmfleft{l1}
   \fmfright{r1}
    \fmftop{t1}
   \fmf{plain}{l1,v1}
    \fmf{plain}{v2,r1}
    \fmf{dashes,right,tension=0.8}{v1,v2}
    \fmffreeze
    \fmfforce{60,82}{v3}
    \fmf{plain,left,tension=0.8}{v1,v2}
    \fmf{wiggly}{v3,t1}
    \fmflabel{$b$}{l1}
    \fmflabel{$s$}{r1}
    \fmflabel{$\gamma$}{t1}
\end{fmfgraph*}}
\end{fmffile} 
&
\hspace{2cm}
&
\begin{fmffile}{diagrams/bsgamma2}
 \fmfframe(0,0)(0,0){
   \begin{fmfgraph*}(120,120)
    \fmfleft{l1}
   \fmfright{r1}
    \fmftop{t1}
   \fmf{plain}{l1,v1}
    \fmf{plain}{v2,r1}
    \fmf{plain,right,tension=0.8}{v1,v2}
    \fmffreeze
    \fmfforce{60,82}{v3}
    \fmf{dashes,left,tension=0.8}{v1,v2}
    \fmf{wiggly}{v3,t1}
    \fmflabel{$b$}{l1}
    \fmflabel{$s$}{r1}
    \fmflabel{$\gamma$}{t1}
\end{fmfgraph*}}
\end{fmffile} 
\end{tabular}
\caption{One loop contributions to $b\to s \gamma$ in the presence of $\RpV$ couplings with a SM fermion $f$ and a sfermion $\tilde{f}$. Either contributions from  $\lambda'_{ij2} \lambda'_{ij3}$, $\lambda'^*_{i2j} \lambda'_{i3j}$ or $\lambda''_{ij2} \lambda''_{ij3}$ are possible.}
\label{fig:bsgamma}
\end{figure}
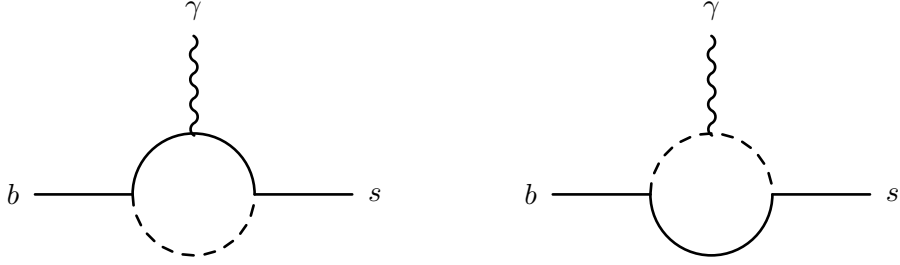
\begin{enumerate}
\item $\sfd{LLE\times LQD}$: Taking into account both trilinear
  $\slashed L$ operators, there are $s$-channel tree level
    decays into $\mu^+\mu^-$, as shown on the left in
  Fig.~\ref{fig:DiagramsProp}. Possible combinations are
\begin{eqnarray}
  \label{eq:lambda1}
  &\lambda_{i22}^*\lambda^\prime_{ij3}\neq 0 \hspace{1cm} \text{or}\hspace{1cm}  \lambda_{i22}^*\lambda^\prime_{i3j} \neq 0 \, , &
\end{eqnarray}
with $i=1,3$ as well as $j=2$ (for $B_s$) or $j=1$ (for $B_d$). Since
$\lambda$ is antisymmetric in its first two indices, the case $i=2$ is
vanishing. However, if one includes other sources of flavor violation
like the $t$--$W$--loop in the SM or possible SUSY loops, 
other combinations of couplings can cause sizable contributions:
\begin{eqnarray}
  \label{eq:lambda1new}
  &\lambda_{i22}^*\lambda^\prime_{ijX}\neq 0 \hspace{1cm} \text{or}\hspace{1cm}  \lambda_{i22}^*\lambda^\prime_{iXj} \neq 0 \, . &
\end{eqnarray}
For the cases $(j,X)\in \left\{ (1,2),(1,1),(2,2),(3,3) \right\}$ a
`SUSY-penguin' is possible with the exchange of a sneutrino
$\spa \nu_i$. We call these cases {\it indirect tree level}.
Also $(j,X)\in \left\{ (3,1),(1,3)\right\}$ cause indirect tree level decays
but these combinations are better constrained by the $B_d^0$ decays.

\item $\sfd{LQD  \times  LQD}$: $t$-channel tree level decays (see Fig.~\ref{fig:DiagramsProp} on the right) are possible with $\sfd L\sfd Q\sfd D$ operators only: 
\begin{eqnarray}
  \label{eq:lambda2}
  &  \lambda^{\prime *}_{2ij}\lambda^\prime_{2i3}\neq 0 \,,  &
\end{eqnarray}
with $j= 2$ (for $B_s$) and $j=1$ (for $B_d$). However, also here it is possible that other loops already change the flavor of the involved quarks leading to indirect tree level decays. This could then cause new contributions for the following pairs of couplings:
\begin{eqnarray}
  \label{eq:lambda2new}
  &  \lambda^{\prime *}_{2ij}\lambda^\prime_{2iX}\neq 0 \,,  &
\end{eqnarray}
with $(j,X)\in \left\{(1,2),(1,1),(2,2),(3,3)\right\}$. Other possible combinations are already covered by eq.~(\ref{eq:lambda2}). 
For $j=X$, it is even possible to constrain not a pair of couplings but single couplings:
\begin{eqnarray}
  \label{eq:lambda2sing}
  &  |\lambda^{\prime}_{2i2}|^2 \neq 0 \,.  &
\end{eqnarray}
One-loop contributions via $Z$ or Higgs penguins (see Fig.~\ref{fig:diagramsLoop}, and for more details Fig.~\ref{fig:Zpeng}) can be triggered by several combinations of $\sfd{LQD}$ couplings. 
\begin{align}
&  \lpa{ijk}{ij3}\neq 0, \\
&  \lpa{ikj}{i3j}\neq 0 
\end{align}
with $k=2$  (for $B_s$) or $k=1$ (for $B_d$). For $k=2$, the same couplings contribute also to $B\to X_s\gamma$ via the diagrams depicted in Fig.~\ref{fig:bsgamma}. 
\item$\sfd{UDD\times UDD}$: If we consider the $\sfd U\sfd D\sfd D$ operator, the products of couplings
\begin{align}
  &\lambda^{\pprime *}_{i12}\lambda^\pprime_{i13} \neq 0\hspace{1cm} (\text{for } B_s), \\
  &\lambda^{\pprime *}_{i21}\lambda^\pprime_{i23} \neq 0\hspace{1cm} (\text{for } B_d),
\end{align}
allow for one-loop decays. The combinations $\lambda^{\pprime *}_{i12}\lambda^\pprime_{i13}$ cause also new contributions to $B\to X_s\gamma$. 
\end{enumerate}
There are, of course, also other combinations of parameters which
could contribute to other decays like those with two electrons, two
$\tau$s or two different lepton flavors in the final state. However,
the experimental limits for these observables are much weaker. In
practice, these parameters just receive upper limits in the case of
tree level decays which has been studied in
Ref.~\cite{Dreiner:2006gu}. Furthermore, pairs of $\lambda$-couplings
can cause new contributions to lepton flavor violating observables
like $\mu \to 3 e$ at tree- and one-loop level. This has already been
studied in Ref.~\cite{Dreiner:2012mx}.  \smallskip Before we turn to
the numerical analysis, we perform a short
analytical discussion of the decays at one-loop. The
$Z$-penguin contributions usually dominate for not too large
$\tan\beta$ and/or not too light CP-odd scalars. 
The corresponding one-loop diagrams shown on the left in
Fig.~\ref{fig:diagramsLoop} consists of vertex corrections as well as
self-energy corrections as depicted in Fig.~\ref{fig:Zpeng}. The only
masses in the loop are those of one SM fermion $m_f$ and of the
sfermions which we assume for the moment to be degenerate with mass
$m_{\tilde{f}}$.  In addition, we neglect squark mixing in this
discussion.  Using the generic results from
Ref.~\cite{Dreiner:2012dh}, we can express the amplitude $A^{RX}$
corresponding to the effective four fermion operators $(\bar{q}_x
\gamma^\mu P_R q_y) (\bar{\mu} \gamma_\mu P_X \mu)$ (with $X=L,R$) as
\begin{align}
16\pi^2 m_Z^2 A^{RX}_{wave}  = & \lambda^{' *}_{ijx} \lambda'_{ijy} B_1(m_f^2,m_{\tilde{f}}^2) Z_q^R Z_l^X\\
16\pi^2 m_Z^2 A^{RX}_{p_1} = & \frac{1}{2} \lambda^{' *}_{ijx} \lambda'_{ijy} \left[-2 Z_f^L B_0(m_f^2,m_f^2) \right. \nonumber \\
 & \left. + 2 C_0(m_{\tilde{f}}^2,m^2_f,m^2_f) (m_f Z_f^R - Z_f^L m_{\tilde{f}}) + Z_f^L C_{00}(m_{\tilde{f}}^2,m_f^2,m_f^2)\right] Z_l^X \\
16\pi^2 m_Z^2 A^{RX}_{p_2} =  & - \frac{1}{2} Z_{\tilde{f}} \lambda^{' *}_{ijx} \lambda'_{ijy} C_{00}(m_f^2,m_{\tilde{f}}^2,m_{\tilde{f}}^2) Z_l^X
\end{align}
 Here, we assumed a {\bf LQD} $\times$ {\bf LQD} contribution, but similar expressions are obtained in the case of {\bf UDD} $\times$ {\bf UDD}. In addition, we parametrized the chiral coupling of the $Z$ to the SM fermion in the loop with $Z_f^{L,R}$ and to the external quark respectively leptons with $Z_q^{L,R}$ and $Z_l^{L,R}$. $Z_{\tilde{f}}$ is the coupling of the $Z$ to the sfermion in the loop. In addition, we neglected all external momenta and masses. Using the analytical expressions for the Passarino-Veltman integrals given in the appendix of Ref.~\cite{Dreiner:2012dh} the sum of all diagrams can be simplified to
\begin{equation}
\label{eq:mfscaling} A^{RX} \propto \lambda^{' *}_{ijx} \lambda'_{ijy} \frac{m_f}{m_{\tilde{f}}^2} Z_l^X\, . 
\end{equation}
Obviously, there is a strong dependence on the mass of the SM fermion
in the loop. Since this is the case as well for the Higgs penguins
which involve Yukawa couplings, one can expect that there is a large
hierarchy between the bounds derived for the different combinations of
$R$pV couplings depending on the generation of SM particles
involved. Furthermore, since down-type squarks only enter together
with neutrinos, their contribution is always completely
negligible.  This is different than $B\to X_s\gamma$
since the Wilson coefficients $C_7$ and $C'_7$ which trigger this
processes don't have the proportionality to the fermion mass in the
loop \cite{deCarlos:1996yh,Kong:2004cp}. Making the same assumption of
vanishing squark flavor mixing, we can express the coefficients as
\begin{align}
\label{eq:A_bsgamma}
C_7^{\prime} &= -Q_d \sum_{i,j=1}^3 \frac{\lpa{i2j}{i3j}}{4\pi}\left( \frac 1{12 m^2_{\spa d_j}}-\frac 1{6m^2_{\spa \nu_i}}  \right), 
\end{align}
Here, we also took the limit $m_{d_j}^2/m_{\spa \nu_i}^2 \to 0$ in
comparison to Ref.~\cite{deCarlos:1996yh,Kong:2004cp}. There is no
dependence on the internal fermion mass left. This reflects also in
the derived limits which are independent of the
involved generation of SM fermions \cite{deCarlos:1996yh}:
\begin{align}
\label{eq:limitBsg1}
|\lambda^{\prime}_{i2j} \lambda^{\prime}_{i3j}| & < 0.09\left[2 \left(\frac{100~\GeV}{m_{\tilde{\nu}_i}}\right)^2 -\left(\frac{100~\GeV}{m_{\tilde{d}_{R,j}}}\right)^2\right]^{-1} \,,\\
|\lambda^{\prime}_{ij2} \lambda^{\prime}_{ij3}| & < 0.035\left[2 \left(\frac{100~\GeV}{m_{\tilde{e}_{L,i}}}\right)^2 -\left(\frac{100~\GeV}{m_{\tilde{d}_{L,j}}}\right)^2\right]^{-1} \,,\\
\label{eq:limitBsg2}
|\lambda^{\prime\prime}_{i2j} \lambda^{\prime\prime}_{i3j}| & < 0.16\left(\frac{m_{\tilde{q}_{R,i}}}{100~\GeV}\right)^2 \,.
\end{align}

%% file: setup.tex
\section{Numerical Setup}
\label{sec:setup}
For our analysis we have generated \SPheno modules by \SARAH for the
two models {\tt MSSMBpV} (MSSM with Baryon number violating $\RpV$
couplings) and {\tt MSSMTriLnV} (MSSM with trilinear Lepton number
violating $\RpV$ couplings) which are part of the public \SARAH
version \cite{sarah}. The \SPheno modules generated by \SARAH provide
Fortran code which allows a precise mass spectrum calculation using
two-loop renormalization group equations (RGEs) and one-loop
corrections to all masses. In addition, it calculates the decay widths
and branching ratios of all Higgs and SUSY particles and calculates
several observables like $l_i\to l_j \gamma$, $\l_i\to 3l_j$ or
$\Delta\rho$ at full one-loop. We are going to use in the following
especially the predictions for $B_{s,d}^0\to \mu \bar{\mu}$ and $B \to
X_s \gamma$ of the code. The calculation of $B_{s,d}^0\to \mu
\bar{\mu}$ in these \SPheno modules has been discussed in detail in
Ref.~\cite{Dreiner:2012dh}, while $B \to X_s \gamma$ is based on the
results of Ref.~\cite{Lunghi:2006hc}. The parameter scans have been
performed with {\tt SSP} \cite{Staub:2011dp}.
\begin{table}[htpb]
\begin{minipage}[c][10cm][t]{0.3\textwidth}
  \begin{tabular}{|>{\centering}m{3cm}|>{\centering\arraybackslash}m{3cm}|} \hline
    parameter & value \\ \hline
    $\tan\beta$ & 42.27 \\
    $\mu$ & 2207.8~GeV \\
    $B_\mu$ & 3.42$\cdot 10^{5}$~GeV${}^2$ \\
    $M_1$ & -831.0~GeV \\
    $M_2$ & 2310.0~GeV \\
    $M_3$ & 1290.0~GeV \\ 
    $(T_u)_{33}$ & -3170.0~GeV\\
    $(T_d)_{33}$ & 198.0~GeV\\
    $(T_e)_{33}$ & 1280.0~GeV\\
\hline
  \end{tabular}
\end{minipage}
\hspace{2cm}
\begin{minipage}[c][10cm][t]{0.3\textwidth}
  \begin{tabular}{|>{\centering}m{3cm}|>{\centering\arraybackslash}m{3cm}|} \hline
    parameter & value~[$\GeV^2$] \\ \hline
    $m_{H_d}^2$       & 9.59$\cdot 10^{6}$ \\
    $m_{H_u}^2$       &-4.87$\cdot 10^{6}$\\
    $(m^2_{Q})_{1,2}$ & 1.17$\cdot 10^{6}$ \\
    $(m^2_{Q})_3$     & 2.44$\cdot 10^{6}$ \\
    $(m^2_{L})_{1,2}$ & 1.18$\cdot 10^{6}$ \\
    $(m^2_{L})_3$     & 1.50$\cdot 10^{6}$ \\
    $(m^2_{d})_{1,2}$ & 9.83$\cdot 10^{6}$ \\
    $(m^2_{d})_3$     & 9.20$\cdot 10^{6}$ \\
    $(m^2_{u})_{1,2}$ & 8.62$\cdot 10^{6}$ \\
    $(m^2_{u})_3$     & 9.43$\cdot 10^{6}$ \\
    $(m^2_{e})_{1,2}$ & 1.12$\cdot 10^{6}$ \\
    $(m^2_{e})_3$     & 8.67$\cdot 10^{6}$ \\ \hline
  \end{tabular}
\end{minipage}
  \caption{Input parameters for point 2342344${}^\prime$  of Ref.~\cite{Cahill-Rowley:2013gca} evaluated at $Q=160~\GeV$.}
  \label{tab:softparameters}
\end{table}
%
\begin{table}[hbt]
  \centering
\begin{minipage}[c][13cm][t]{0.49\textwidth}
  \begin{tabular}{|>{\centering}m{3cm}|>{\centering}m{2cm}|>{\centering\arraybackslash}m{2cm}|} \hline 
PDG code & Mass~[$\GeV$] & particle \\ \hline
        25 &     124.8   & $h$ \\
        35 &    3724.3   & $H$ \\
        36 &    3724.1   & $A$\\
        37 &    3725.7   & $H^\pm$\\
   1000001 &     925.7   & $\spa d_L$\\
   2000001 &     906.3   & $\spa d_R$\\
   1000002 &     922.5   & $\spa u_L$\\
   2000002 &     889.5   & $\spa u_R$\\
   1000003 &     925.7   & $\spa s_L$\\
   2000003 &     906.4   & $\spa s_R$\\
   1000004 &     922.5   & $\spa c_L$\\
   2000004 &     889.5   & $\spa c_R$\\
   1000005 &    1186.1   & $\spa b_1$\\
   2000005 &    3088.6   & $\spa b_2$\\
   1000006 &    1180.7   & $\spa t_1$\\
   2000006 &    3190.7   & $\spa t_2$\\
 \hline
  \end{tabular}
\end{minipage}
\begin{minipage}[c][13cm][t]{0.49\textwidth}
  \begin{tabular}{|>{\centering}m{3cm}|>{\centering}m{2cm}|>{\centering\arraybackslash}m{2cm}|} \hline 
PDG code & Mass~[$\GeV$] & particle \\ \hline
   1000011 &    1021.8   & $\spa e_L$\\
   2000011 &    1001.8   & $\spa e_R$\\
   1000012 &    1018.5   & $\spa \nu_{eL}$\\
   1000013 &    1021.8   & $\spa \mu _L$\\
   2000013 &    1001.8   & $\spa \mu_R$\\
   1000014 &    1018.5   & $\spa \nu_{\mu L}$\\
   1000015 &     948.2   & $\spa \tau_1$\\
   2000015 &    1074.5   & $\spa \tau_2$\\
   1000016 &    1019.5   & $\spa \nu_{\tau L}$\\
   1000021 &    1248.7   & $\spa g$\\
   1000022 &     853.3   & $\chi_1$\\
   1000023 &    1877.9   & $\chi_2$\\
   1000025 &    1887.2   & $\chi_3$\\
   1000035 &    2317.6   & $\chi_4$\\
   1000024 &    1861.6   & $\chi_1^\pm$\\
   1000037 &    2299.4   & $\chi_2^\pm$ \\\hline
  \end{tabular}
\end{minipage}
  \caption[SUSY spectrum]{SUSY spectrum of benchmark point 2342344${}^\prime$ of Ref.~\cite{Cahill-Rowley:2013gca}.}
  \label{tab:spectrum}
\end{table}
\begin{table}[hbt]
\centering
\small
\begin{tabular}{|l|l|l|l|l|}
\hline \hline
\multicolumn{5}{|c|}{default SM input parameters} \\
\hline
$\alpha^{-1}_{em}(M_Z) = 127.93  $ & $\alpha_s(M_Z) = 0.1190  $ & $G_F = 1.16639\cdot 10^{-5}~\text{GeV}^{-2}$
 & $\rho = 0.135$ & $\eta = 0.349$  \\
$m_t^{pole} = 172.90~\text{GeV}$ & $M_Z^{pole} = 91.1876~\text{GeV}  $ & $m_b(m_b) = 4.2~\text{GeV} $ & $\lambda = 0.2257 $ & $A = 0.814$ \\
\hline \hline
\multicolumn{5}{|c|}{derived parameters} \\
\hline
 $m_t^{\overline{DR}} = 166.4~\text{GeV}$ &
$| V_{tb}^* V_{ts}| = 4.06*10^{-2}  $ & $| V_{tb}^* V_{td}| = 8.12*10^{-3} $ & $m_W = 80.3893 $ & $\sin^2 \Theta_W = 0.2228 $ \\
\hline
\end{tabular}
\caption{SM input values and derived parameters used for the numerical evaluation of \mbox{$\Bsdll$} in \SPheno.} 
  \label{tab:sm}
\end{table}
\begin{table}[hbt]
\centering
\begin{tabular}{|l|l|l|}
\hline
\multicolumn{3}{|c|}{Default hadronic parameters} \\
\hline
$m_{\Bs} = 5.36677$ GeV & $f_{\Bs} = 227(8)$ MeV & $\tau_{\Bs} = 1.466(31)$ ps \\
$m_{\Bd} = 5.27958$ GeV  & $f_{\Bd} = 190(8)$ MeV & $\tau_{\Bd} = 1.519(7)$ ps  \\
\hline
 \end{tabular}
 \caption{Hadronic input parameters used for the numerical evaluation of \mbox{$\Bsdll$} in \SPheno.} 
  \label{tab:input}
\end{table}

As a starting point we choose a point in the MSSM parameter
space, which reproduces the right Higgs mass and is in no conflict
with any other experimental measurements.  A set of such benchmark
points fulfilling these constraints has been proposed in
\cite{Cahill-Rowley:2013gca} within the framework of the
phenomenological MSSM (pMSSM) without \RpV \cite{Berger:2008cq}. This
simplified model consists of a subset of 19 MSSM
parameters in contrast to the most general MSSM with more than 100
parameters. The basic assumptions are CP conservation, Minimal Flavor
Violation, degeneracy of the first and second sfermion generations and
vanishing Yukawa couplings for the first two generations, and the
lightest neutralino as a dark matter candidate. The
benchmark points are chosen to satisfy the latest LHC 7/8~\TeV{}
searches, the Higgs mass at 126~GeV, as well as precision observables
($b\to s \gamma$,\,$(g-2)_\mu$,\,$B_s\to \mu^+\mu^-$,\,$B\to\tau\nu$,
etc.) and cosmological bounds. We chose the point \#~$2342344^\prime$
because it features a compressed and relatively light spectrum
compared to the spectra of other benchmark points. Hence, it is
expected to provide the most significant limits in the following.  In
addition we checked with {\tt Vevacious} that it has a stable,
electroweak vacuum \cite{Camargo-Molina:2013qva}. The input parameters
are given in Tab.~\ref{tab:softparameters} and the spectrum is given
in Tab.~\ref{tab:spectrum}. As SM input and for the hadronic variables
we used the values given in Tab.~\ref{tab:sm} and \ref{tab:input}.

The MSSM predicitions without $R$pV for this point for the $B$-decays are a little bit above the SM predictions, but well inside the allowed range. 
\begin{equation}
\text{BR}(B_s^0\to \bar{\mu}\mu) = 3.86\cdot 10^{-9} \, , \hspace{1cm} \text{BR}(B_d^0\to \bar{\mu}\mu) = 1.26\cdot 10^{-10} \, .
\end{equation}
The origin of the difference is a significant contribution stemming
from chargino loops.  To study the impact of the $R$pV couplings, we
used the running parameters for this point at $Q=160$~GeV as
calculated by \SPheno in the MSSM. In this way we disentangle the
effect of the new parameters in the RGE evolution. Afterwards $\RpV$
is \enquote{switched on} by raising the values of certain combinations
of \RpV couplings, which enhance or reduce the branching ratios. For
each combination of couplings, one is kept fixed and the other one is
varied within a range from $10^{-5}$ up to $\mathcal{O}(4\pi)$.

It is important to mention that the product of two couplings (say,
$\lpa{ijk}{lmn}$) has a relative phase $\sigma$ to the SM or other
SUSY contributions, which can be chosen freely. To study the impact of
this phase on the branching ratios, we use both, $\lambda^{\prime
  *}\lambda^\prime>0$ and $\lambda^{\prime *}\lambda^\prime<0$, in the
scans. The same holds for the other 
$R$pV couplings. Usually, there is constructive interference with the
SM contributions for one sign and destructive interference for the
other sign. We concentrate in our studies on the case of real
$\lambda$-couplings. The impact of complex $R$pV couplings on
tree level decays is discussed in Ref.~\cite{Yeghiyan:2013upa}.

%% file: results.tex
\section{Results}
\label{sec:results}
\subsection{Tree Level Results}
\subsubsection{Direct Tree Level}
\begin{figure}[htpb]
  \centering
    \includegraphics[width=0.49\textwidth]{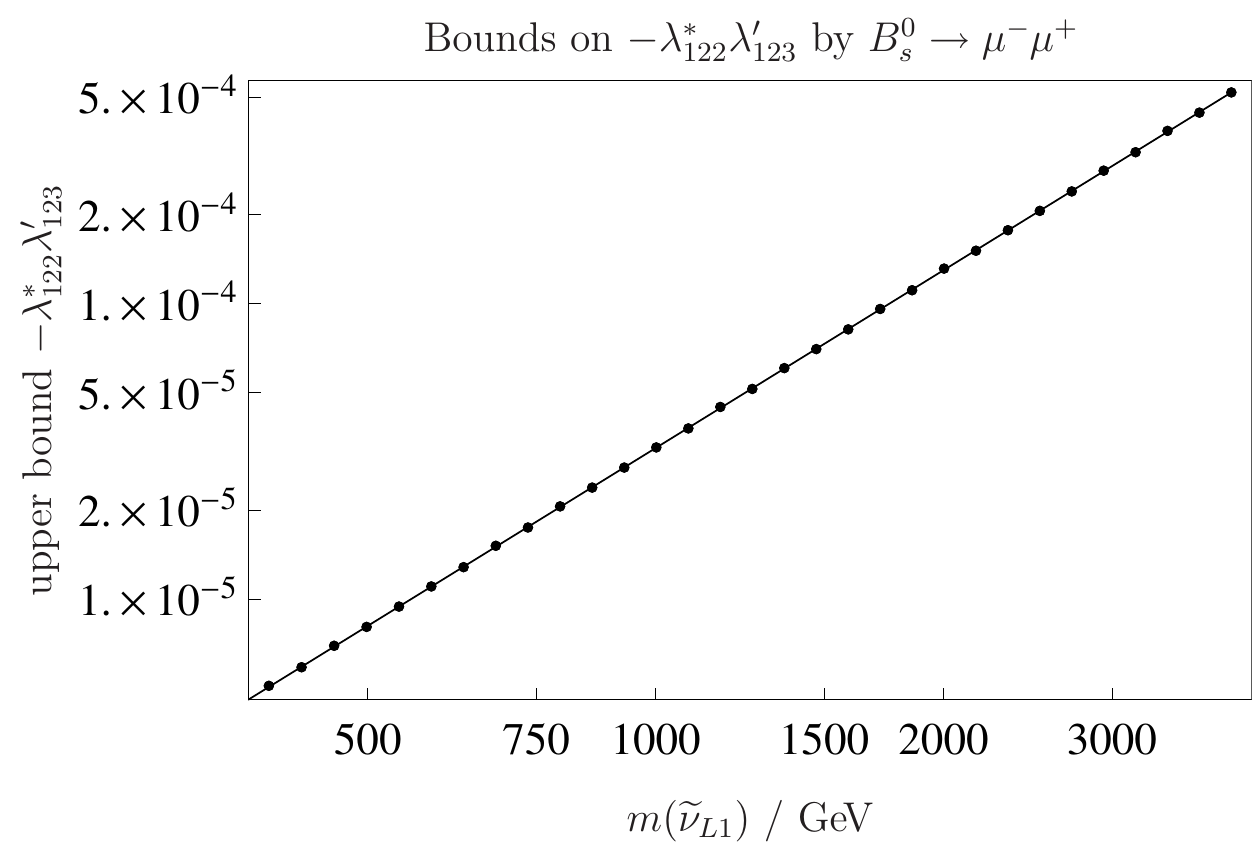} \hfill 
    \includegraphics[width=0.49\textwidth]{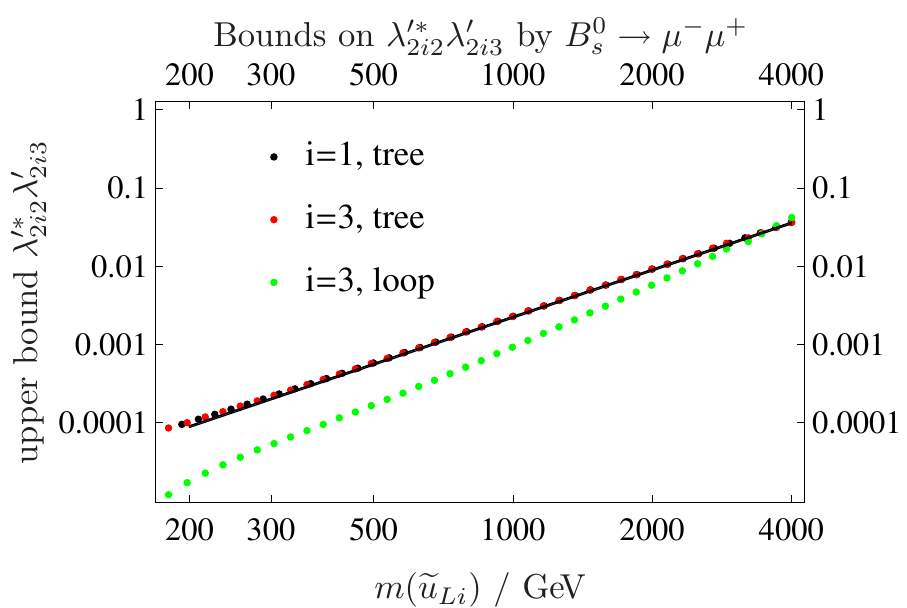} 
  \caption{Limits on $R$pV couplings as function of the sfermion mass in the propagator. Left: Bounds for $-\lambda_{122}\lambda^\prime_{123}$ ($s$-channel). Right: Bounds for $\lpa{2i2}{2i3}$ ($t$-channel). The black dots ($i=1$) and red dots ($i=3$) follow the $\sim {\tilde m^2}$ scaling. They only contain tree level contributions and SM one-loop contributions. The green dots include the full one-loop SUSY contributions and do not follow this scaling. }
  \label{fig:treelevelLQD}
\end{figure}

\begin{table}[t]
  \centering
   \begin{tabular}{|p{5cm}|>{\centering}m{4cm}|>{\centering\arraybackslash}m{4cm}|} \hline
  \multicolumn{3}{|c|}{Tree Level} \\
  \hline
    $\lambda^*\lambda^\prime$ & $\lambda^*\lambda^\prime > 0$ &  $\lambda^*\lambda^\prime < 0$  \\
    $i22\ i23\ (\text{and }i32),\ i=1,2$ & $< 2.51 \times 10^{-10} [m_{\spa \nu_i}^2] $ & $>-3.25 \times 10^{-11} [m_{\spa \nu_i}^2]$ \\
    $i22\ i23\ (\text{and }i32),\ i=3$ & $< 2.31 \times 10^{-10} [m_{\spa \nu_i}^2] $ & $>-3.25 \times 10^{-11} [m_{\spa \nu_i}^2]$ \\
    $i22\ i13$, $i=1,2,3$ & $< 7.06 \times 10^{-11} [m_{\spa \nu_i}^2]$ &  $>-1.15 \times 10^{-10} [m_{\spa \nu_i}^2]$ \\
    $i22\ i31$, $i=1,2,3$ & $< 7.06 \times 10^{-11} [m_{\spa \nu_i}^2]$ &  $>-1.15 \times 10^{-10} [m_{\spa \nu_i}^2]$ \\
    \hline
    $\lambda^{\prime*}\lambda^\prime$ & $\lambda^{\prime*}\lambda^\prime >0$ & $\lambda^{\prime*}\lambda^\prime <0$ \\
    $2i2\ 2i3$, $i=1,2$ & $< 2.26 \times 10^{-9} [m_{\spa u_{Li}}^2]$ & $>-4.68 \times 10^{-9}[m_{\spa u_{Li}}^2]$  \\
    $2i2\ 2i3$, $i=3$ &  $< 2.25 \times 10^{-9} [m_{\spa u_{Li}}^2]$ & $>-4.66 \times 10^{-9}[m_{\spa u_{Li}}^2]$  \\
    $2i1\ 2i3$, $i=1,2$ &$< 1.20 \times 10^{-8}[m_{\spa u_{Li}}^2] $ & $ >-5.64 \times 10^{-9}[m_{\spa u_{Li}}^2]$ \\
    $2i1\ 2i3$, $i=3$ & $< 1.13 \times 10^{-8} [m_{\spa u_{Li}}^2]$ & $ >-5.62 \times 10^{-9}[m_{\spa u_{Li}}^2]$ \\
    \hline
  \end{tabular}
    \caption{Collection of bounds from decays at tree level on pairs of $R$pV couplings. We used the parameter point given in Tab.~\ref{tab:spectrum} based  on Tab.~\ref{tab:softparameters}. The notation $[m_{\spa f}^2]$ means $m_{\spa f}^2/(\GeV)^2$. We neglected here other SUSY contributions beside the $R$pV ones.}
  \label{tab:collection}
\end{table}
We first present our results for the only case which has been so far considered in the literature for semi-leptonic $B$-meson decays in the context of broken $R$-parity: combinations of couplings which can cause these decays at tree level. In general the tree level diagrams are not the only relevant SUSY contributions. Especially chargino loops can have a non negligible effect as shown in the previous section. However, we ignore these contributions as is usually done in the literature for a moment since they introduce a dependence on several SUSY masses and parameters.  Under this assumption, the matrix element is proportional to $\lambda^*\lambda^\prime/m_{\spa {f} }^2$, i.e. it depends only on the $R$pV couplings and the mass of the propagating sfermion.  This scaling is also reflected in our numerical analysis as shown for a representative case on the left in Fig.~\ref{fig:treelevelLQD}. Our entire results are summarized in Tab.~\ref{tab:collection}. We pick only one representative case for the following discussion and comparison with previous studies. The extracted limits for $\lambda_{i22} \lambda'_{i23}$ are \begin{subequations} \label{eq:lambdabounds1} \begin{align}
 \label{eq:lambdabounds1a} -3.25 \times 10^{-11} [m_{\spa \nu_i}]^{2} &< \lambda_{i22} \lambda'_{i23}  < 2.51 \times 10^{-10} [m_{\spa \nu_i}]^{2} \, ,\hspace{1cm} i=1,2,\ \text{same for}\ \lambda_{i22}\lambda'_{i32}\\
 \label{eq:lambdabounds1b} -1.15 \times 10^{-10} [m_{\spa \nu_i}]^{2} &< \lambda_{i22}\lambda'_{i13}  < 7.06 \times 10^{-11} [m_{\spa \nu_i}]^{2} \, ,\hspace{1cm} i=1,2,3,\ \text{same for}\ \lambda_{i22}\lambda'_{i31}. 
 \end{align}
\end{subequations}
Recently published results for these bounds have been \cite{Li:2013fa}
\begin{equation}
\label{eq:lambdaboundsRef}
\abs{\sum_i \frac{\lambda_{i22}\lambda'_{i23}}{[m_{\spa \nu_i}]^{2}}}  < 6.52\cdot 10^{-11} \,, \hspace{1cm} \abs{\sum_i \frac{\lambda_{i21}\lambda'_{i23}}{[m_{\spa \nu_i}]^{2}}}  < 7.85\cdot 10^{-11}.
\end{equation}
As we pointed out in the introduction, a product like $\lambda^*\lambda^\prime$ (or $\lambda^{\prime *}\lambda^\prime$) has a phase $\sigma$ relative to the SM contributions. By choosing $\sigma=\pm 1$, we obtain the positive and negative bound given in eqs.~(\ref{eq:lambdabounds1}). For $\sigma=-1$ constructive interference between the SM and $R$pV contributions appears and we obtain the stronger limit ($-3.25\times 10^{-11} [m_{\spa \nu_i}]^{2}$), which is comparable to the one in eq.~(\ref{eq:lambdaboundsRef}) from Ref.~\cite{Li:2013fa}, but slightly better, because we used an updated experimental bound.  
The reason that the limit for the destructive phase is much weaker is not only the asymmetric bounds in eq.~(\ref{eq:ranges}), but also based on the different operators which enter the calculation. The $\RpV$ contributions stem only from sneutrino exchange diagrams at tree level, which contribute to the scalar and pseudoscalar coefficients $F_S,F_P$ introduced in eq.~(\ref{eq:squaredMBsllp}). While $F_S^{RpV} = - F_P^{RpV}$ holds, $\Bsmm$ in the SM is determined mainly by the axial vector coefficient $F_A$. 

Also for $\lambda' \lambda'$ we find the expected scaling if one neglects other SUSY contributions, see Fig.~\ref{fig:treelevelLQD} (right).
To demonstrate the effect of the other SUSY loops on this scaling we compare on the right in Fig.~\ref{fig:treelevelLQD} the case with and without the other SUSY contributions. One can see that especially for light sfermions this can cause a pronounced difference and leads not only to an off-set but also to a different slope. Hence, if one studies areas in the parameter space of $R$pV SUSY containing light squarks it might not be sufficient to consider just the simplified limits usually discussed in this context, but each point has to be studied carefully. 

\subsubsection{Indirect Tree Level Results}
\begin{table}[t]
  \centering
 \begin{tabular}{|p{4cm}|>{\centering}m{4cm}|>{\centering\arraybackslash}m{4cm}|} \hline
    \multicolumn{3}{|c|}{Indirect Tree Level} \\
    \hline
    $\lambda^{\prime *}\lambda^\prime$ & {$\lambda^{*\prime}\lambda^\prime > 0$} &  {$\lambda^{*\prime}\lambda^\prime < 0$}  \\
     $2i2\ 2i2$, $i=1,2$ &  {$<8.2\times 10^{-5}[m_{\spa u_i}^2] $} & - \\
$2i3\ 2i3$, $i=1,2$ &  {$<1.5\times 10^{-5} [m_{\spa u_i}^2]$} & - \\
     $2i2\ 2i2$, $i=3$ &  {$<3.1\times 10^{-7} [m_{\spa u_i}^2]$}  &  - \\
$2i3\ 2i3$, $i=3$ &  {$<6.0\times 10^{-5} [m_{\spa u_i}^2]$} &  - \\ 
	\hline
    $\lambda^* \lambda^\prime$ &  {$\lambda^{*}\lambda^\prime > 0$} &  {$\lambda^{*}\lambda^\prime< 0$}  \\ \hline
    $i22\ i12$, $i=1,2,3$ &  {$< 3.3 \times 10^{-7} [m_{\spa \nu_i}^2]$} &  {$>-2.6 \times 10^{-8} [m_{\spa \nu_i}^2]$} \\  
    $i22\ i22$, $i=1,2,3$ &  {$<5.1 \times 10^{-9} [m_{\spa \nu_i}^2]$} &  {$>-7.9 \times 10^{-8} [m_{\spa \nu_i}^2]$} \\
    $i22\ i33$, $i=1,2,3$ &  {$<6.8\times 10^{-8} [m_{\spa \nu_i}^2]$} &  {$>-4.0 \times 10^{-9} [m_{\spa \nu_i}^2]$} \\
    $i22\ i21$, $i=1,2,3$ &  {$<3.8 \times 10^{-8} [m_{\spa \nu_i}^2]$} &  {$>-2.2\times 10^{-8} [m_{\spa \nu_i}^2]$} \\
    $i22\ i11$, $i=1,2,3$ &  {$<1.0 \times 10^{-7} [m_{\spa \nu_i}^2]$} &  {$>-1.8 \times 10^{-7} [m_{\spa \nu_i}^2]$} \\
    \hline
   \end{tabular}
   \caption{Collection of bounds from decays with sfermions in the propagator which rely on a flavor change in one SM or SUSY loop ('indirect tree level').  We used the input parameters given in Tab.~\ref{tab:softparameters}. The notation $[m_{\spa f}^2]$ means $m_{\spa f}^2/(\GeV)^2$.}
  \label{tab:indirect_tree}
\end{table}
There is another class of combinations for SUSY-penguins described in eq.~(\ref{eq:lambda1new}) which rely on an additional flavor change in either the SM or another SUSY loop. The results for these couplings are given in Tab.~\ref{tab:indirect_tree}. Here, we have included all SUSY corrections present in our benchmark scenario. While the bounds for $\lambda^* \lambda^\prime$ scale proportional to  $m_{\spa \nu_i}^2$ to an high accuracy, this holds for the dependence of the limits for $\lambda^{*\prime}\lambda^\prime$ on $m_{\spa u_i}^2$ only to some extent: the squark masses appear not only in the propagator but also in the important chargino loop. 
In general the resulting limits are worse than for the pure tree level contributions: 
for our benchmark point with $m_{\spa \nu_i}\approx 1~\TeV$, the bounds are between $0.33$ and $4\times 10^{-3}$. Nevertheless, these limits are still competitive with those of {\it direct} tree level semi-leptonic  decays of other mesons. For instance, the combination $\lambda_{i22}^*\lambda^{\prime}_{i12}$ is also constrained by searches for $K^0\to \mu \bar \mu$. The tree level limit based on this observable is given by \cite{Barbier:2004ez}  
\begin{equation}
\abs{\lambda^*_{i22} \lambda^\prime_{i12}} < 2.2\times 10^{-7} [m_{\spa \nu_L}^2] \, ,
\end{equation}
which we improve on at one side of our asymmetric bound by about one order magnitude.

We don't give the limits for other parameter combinations of $\lambda^{\prime *} \lambda^{\prime}$ which would in principle also contribute to indirect tree level decays: $(2ij)\times (2i3)$ and $(2ij)\times (2i2)$ ($j\neq 2$). These are partially already constrained by $B^0_d\to \mu\bar{\mu}$. In addition, there is always a superposition of two contributions $|\lambda_{2i3}|^2+\lambda^{\prime *}_{2ij}\lambda_{2i3}$ which does not provide a clean environment to derive limits only on just one combination. 

\subsection{One-Loop Results}
\begin{table}[t]
  \centering
\begin{tabular}{|p{3cm}|>{\centering}m{3cm}|>{\centering}m{3cm}|>{\centering}m{3cm}|>{\centering\arraybackslash}m{3cm}|}  
   \hline
    \multicolumn{5}{|c|}{One-Loop Level} \\ \hline
    \hline
    $\lambda^{\prime*}\lambda^\prime$  &  \multicolumn{2}{c|}{$\lambda^{\prime*}\lambda^\prime > 0$} & \multicolumn{2}{c|}{$\lambda^{\prime*}\lambda^\prime< 0$}  \\ 
    & $B_{s,d}\to\bar\mu\mu$ & \bsgamma & $B_{s,d}\to\bar\mu\mu$ & \bsgamma \\ 
    \hline
    $3j2\ 3j3,\ j=1,2$  & $< 2.89$ & $<8.80$ & $\O{}$& $>-5.45 $  \\
    $3j2\ 3j3$, $j=3$ & $< 0.49$ & $<2.57$ & $>-0.085$ & $>-6.24$ \\
    $32j\ 33j,\ j=1,2$  & $\O{}$ & $<5.28$ & $\O{}$ &  $>-1.15$ \\
    $32j\ 33j$, $j=3$ & $<11.25$ & $<5.28$ & $\O{}$ & $>-1.15$ \\
    $331\ 333$ & $< 0.45$ & $X$ &  $>-0.96 $ & $X$  \\
    $i1j\ i3j$ & $\O{}$  & $X$& $\O{}$ & $X$ \\
    \hline 
    $\lambda^{\pprime*}\lambda^\pprime$ &   \multicolumn{2}{|c|}{$\lambda^{\pprime*}\lambda^\pprime >0$} &  \multicolumn{2}{|c|}{$\lambda^{\pprime*}\lambda^\pprime<0$} \\ 
    \hline
    $i12\ i13,\ i=1,2$ & $\O{}$& $<0.69$ & $\O{}$ & $>-2.46$ \\
    $312\ 313$ & $< 0.178 $ & $<5.00$& $>-0.030$& $>-6.00$ \\
    $i21\ i23,\ i=1,2$ & $\O{}$& $X$ & $\O{}$ & $X$ \\
    $321\ 323$ & $<0.162$ & $X$ & $>-0.347$ & $X$\\ \hline 
  \end{tabular}
\caption{Collection of bounds for decays at one-loop level for the input parameters given in Tab.~\ref{tab:softparameters}. $\O{}$ means that the limit is outside the perturbative range, while the couplings marked with $X$ don't contribute to $\bsgamma$. The parameter dependence of the bounds is discussed in detail in the text.}
  \label{tab:collection2}
\end{table}
In Tab.~\ref{tab:collection2} we summarize the limits on those RpV couplings which lead to additional one-loop contributions to the neutral $B$-meson decays. In general one can see that in most cases, in which it is possible at all to obtain a limit from the leptonic decays, that the limits are stronger than for the radiative decay. The is the case for couplings which include heavy SM fermions. In contrast $\bsgamma$ puts limits on all combination independently of the involved SM fermion of roughly the same order. This different behavior can be understood from eq.~(\ref{eq:mfscaling}) in comparison to eq.~(\ref{eq:A_bsgamma}). The obtained limits for $\bsgamma$ are in agreement with previous results given in eqs.~(\ref{eq:limitBsg1}) to (\ref{eq:limitBsg2}) but slightly stronger since we included the chargino loops in our analysis.

Since it is not possible to parametrize the limits as a function of the relevant SUSY masses in contrast to the tree level decays, we are going to discuss the dependence on the different masses and parameters in more detail in the following.

\subsubsection{One-Loop Results for {\bf LQD}}
\label{sec:resultsLQD}
\begin{figure}[tb]
\centering
(a) \includegraphics[width=0.45\linewidth]{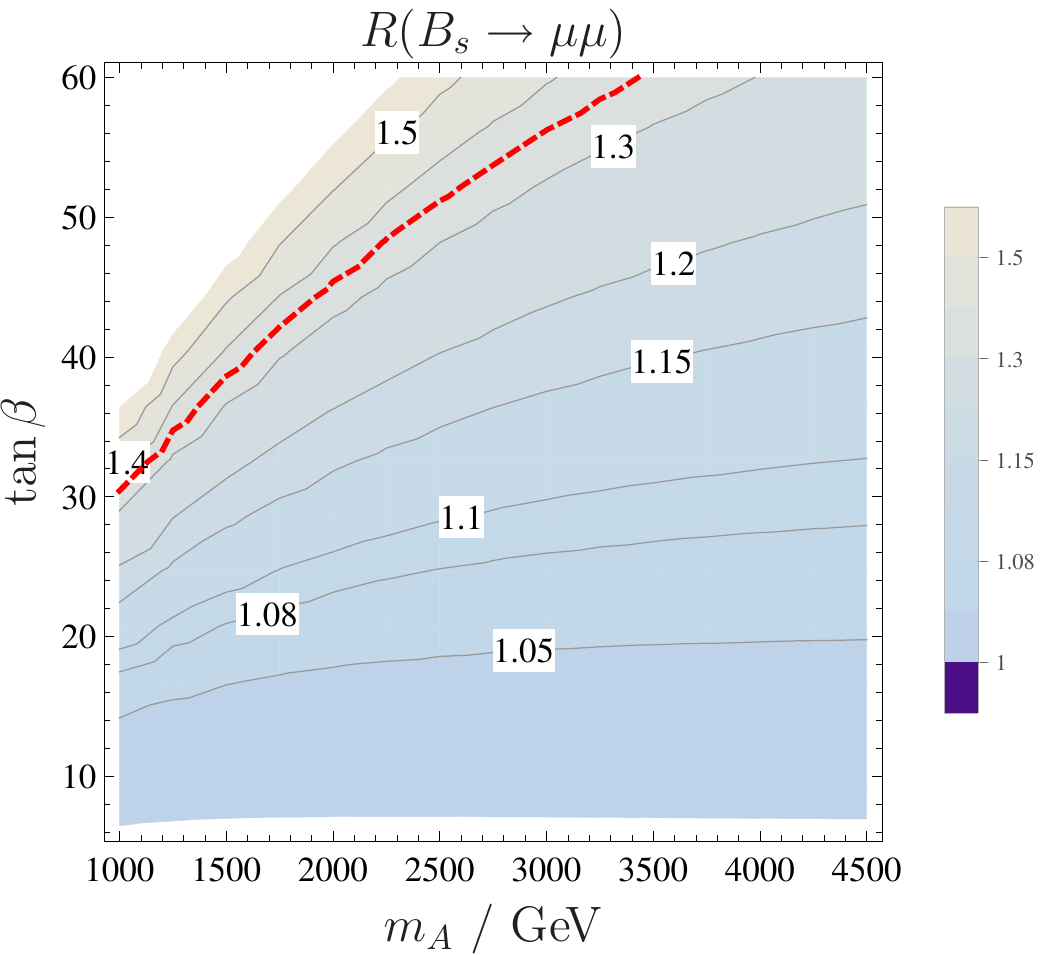} 
(b) \includegraphics[width=0.45\linewidth]{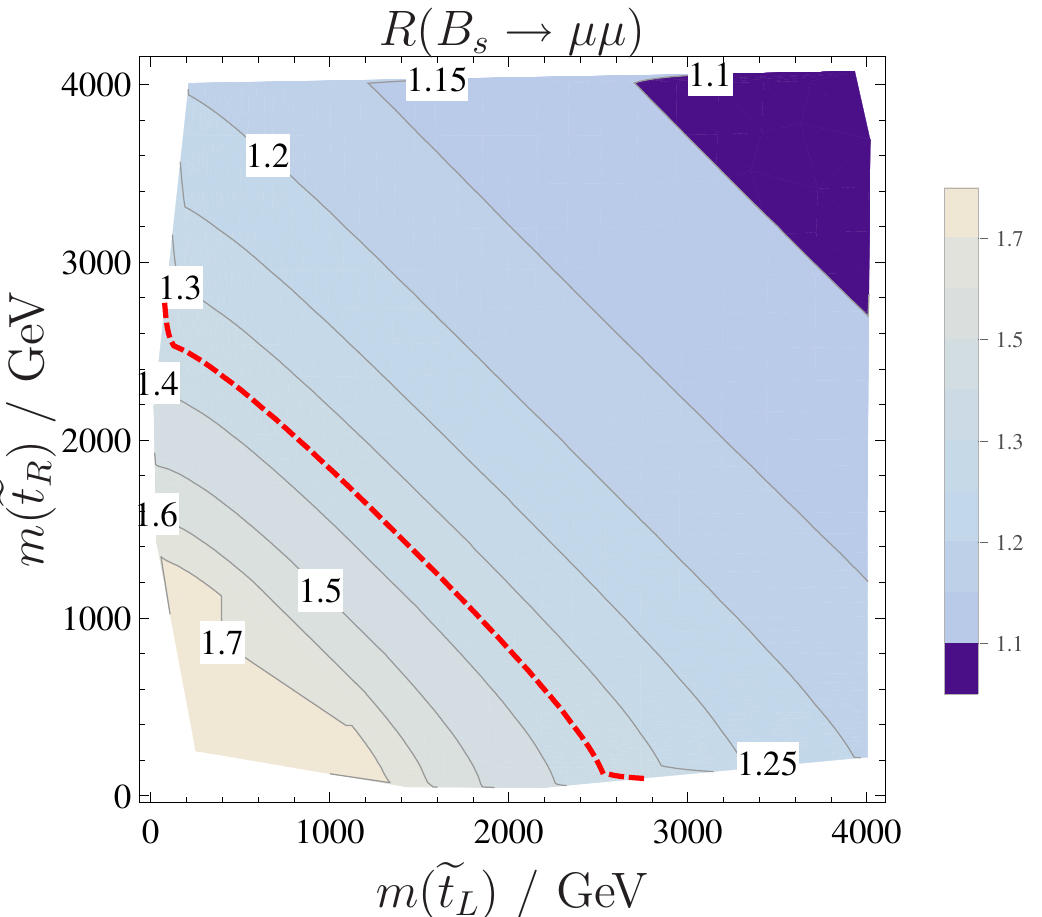} 
\\
(c) \includegraphics[width=0.45\linewidth]{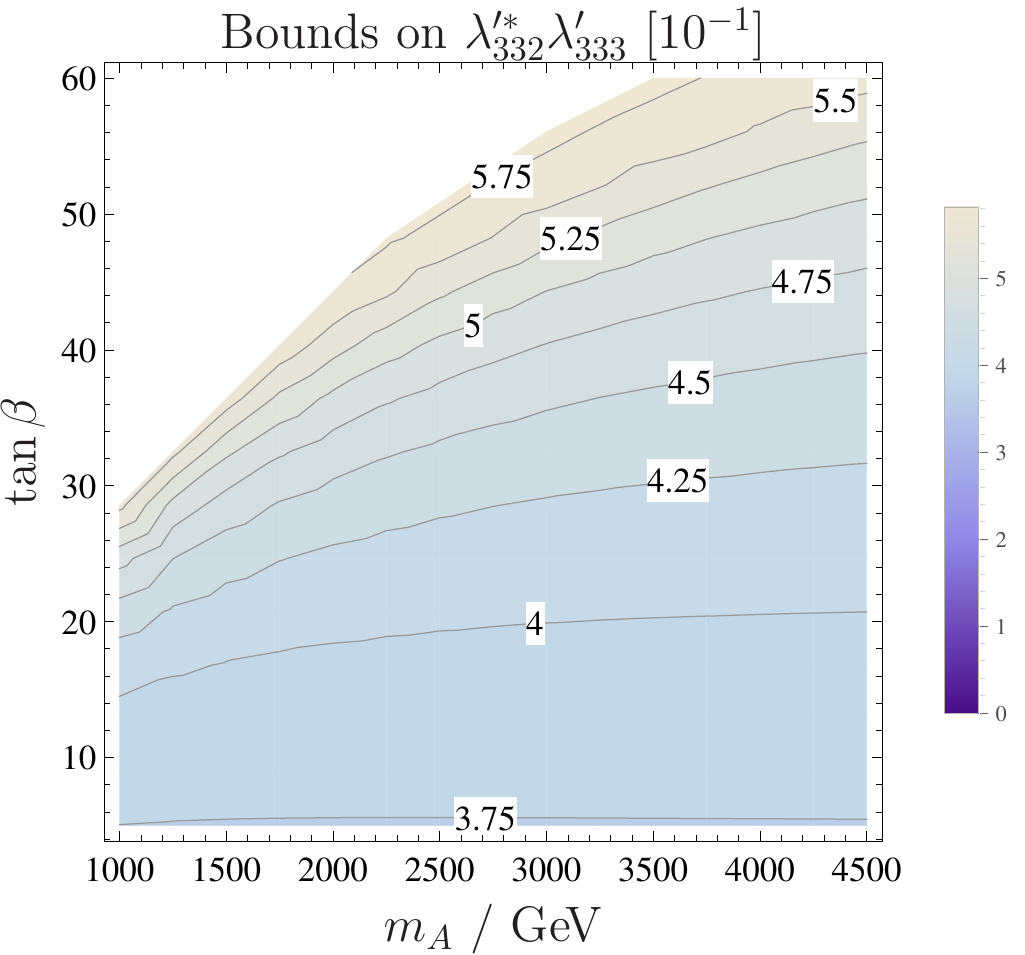}
(d) \includegraphics[width=0.45\linewidth]{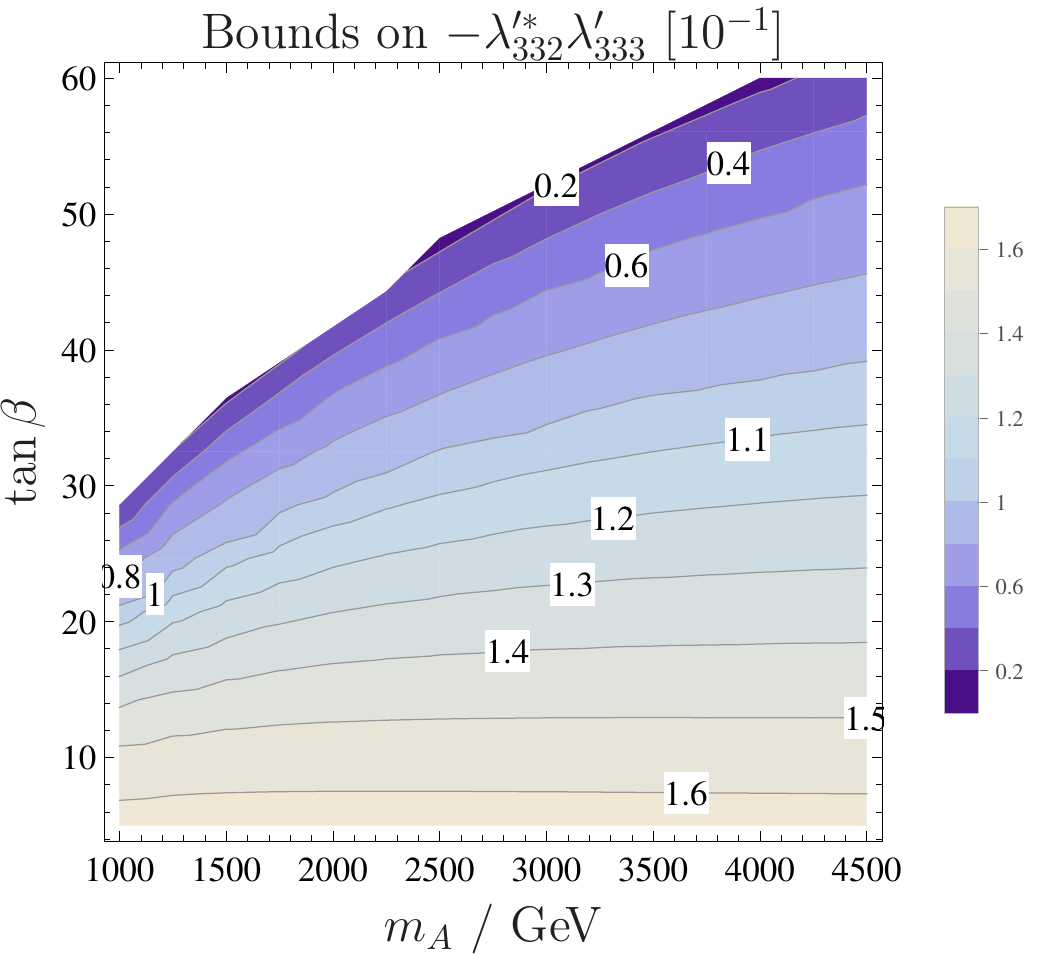}
\caption{First row: $R_s$ is plotted in the $m_A$--$\tan\beta$ plane (left, exclusion above the red dashed line) and in the $m_{\spa t_{L/R}}$ plane (right, exclusion below the red dashed line) without $R$pV contributions. Second row: The left picture shows the upper bounds on $\lpa{332}{333}>0$ in the $m_A$--$\tan\beta$ plane, the right picture shows the upper bound on $\lpa{332}{333}<0$. All other parameters are fixed to Tab.~\ref{tab:softparameters}. Note that the bounds are scaled by a factor of $10^{-1}$.}
\label{fig:MATanbScan}
\end{figure}
It is well known that in the MSSM the most important SUSY corrections to $B_s^0\to\mu\mu$ are due to chargino-loops \cite{bsmumu-susy}. These loops have a very strong dependence on $\tan\beta$ and scale as $\propto \tan^6\beta$ \cite{Buras:2002vd}.  Therefore, we start with a discussion of this effect and check how the bounds for our benchmark point change as a function of $\tan\beta$. The results for each scan are given in a contour plot where the height corresponds to the upper bound of the $\lambda^{*\prime} \lambda^\prime$ combination. Note, the contours are rescaled by a factor given in the title of the plot. The variation of the different parameters changes, of course, also the Higgs mass. However, the contribution of a SM-like Higgs to the observables under consideration is negligible. Therefore, it is not necessary to take this effect into account in the following discussions. 
\begin{figure}[tpb]
  \centering
   \includegraphics[width=0.49\textwidth]{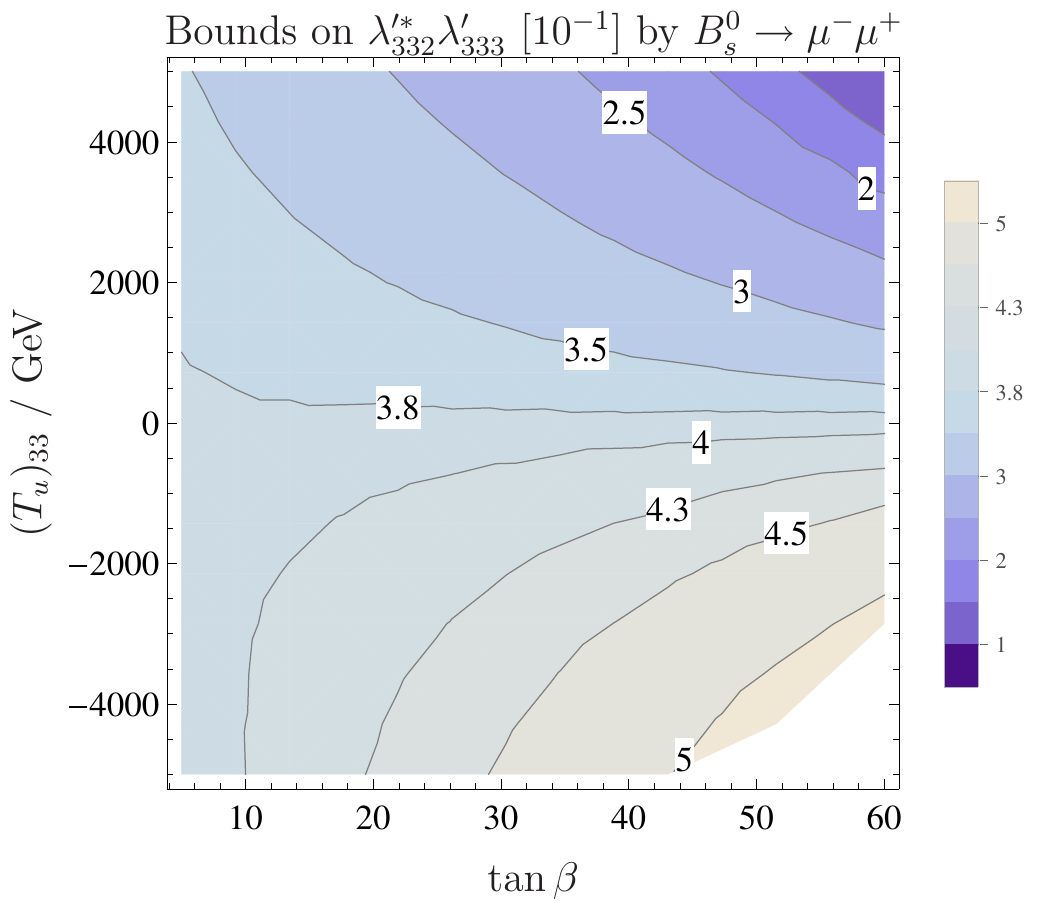} 
   \includegraphics[width=0.49\textwidth]{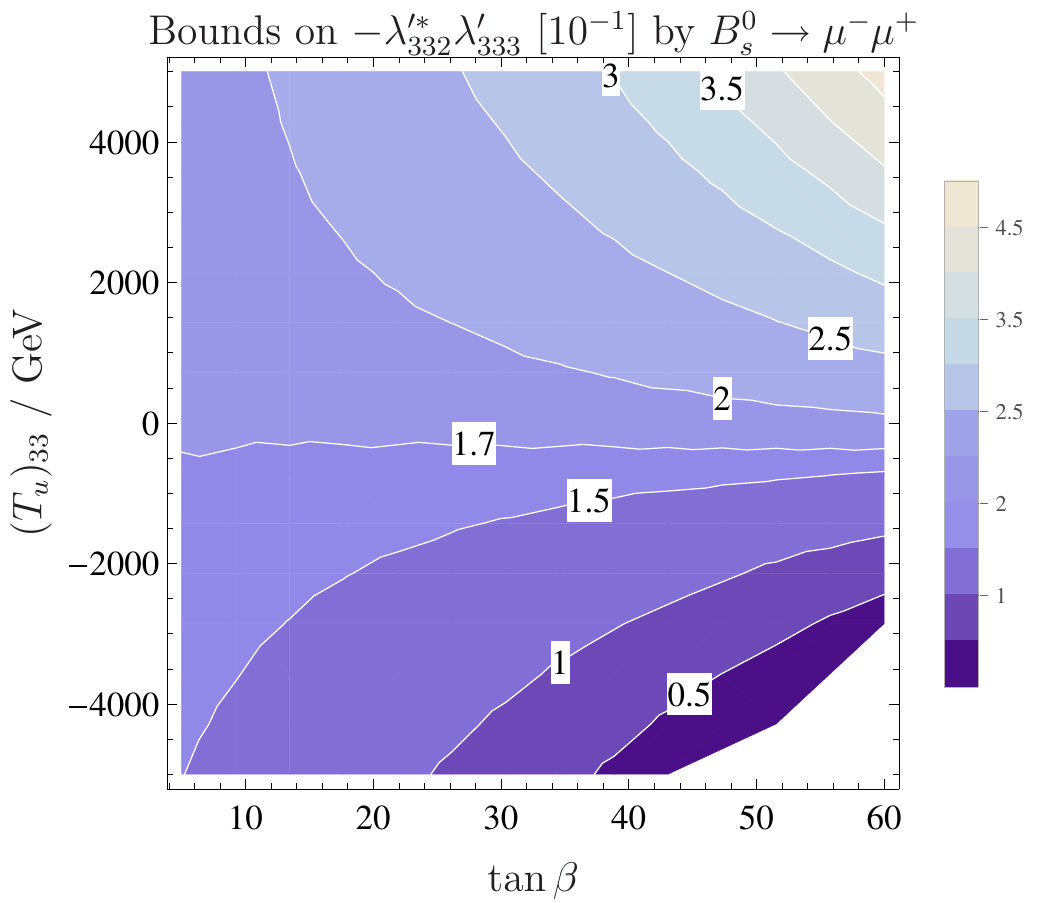} \hfill
  \caption{Upper bounds on $\lpa{332}{333}$ for a variation of $\tanb$ and $(T_u)_{33}$. On the left we assumed $\lpa{332}{333}>0$ and on the right $\lpa{332}{333}<0$. The other parameters are fixed to Tab.~\ref{tab:softparameters}.}
  \label{fig:oneloopLQDtanbTu}
\end{figure}

The plots in the first row of Fig.~\ref{fig:MATanbScan} show an exclusion limit for large $\tan\beta$ with small $m_A$ (Fig.~\ref{fig:MATanbScan}a, above the red dashed line) and for small stop masses $m_{\spa t_{L/R}}$ (Fig.~\ref{fig:MATanbScan}b, below the red dashed line) in the absence of any $R$pV contributions. Based on this the bounds on \RpV couplings derived have a strong indirect dependence on $\tan\beta$. While the $R$pV contributions themselve have only a very small dependence on $\tan\beta$, the enhancement of the other SUSY loops can change the limits signficantly: in the case of contructive interference (Fig.~\ref{fig:MATanbScan}d) the bounds improve by about one order of magnitude between $\tan\beta=5$ and 50. For destructive interference (Fig.~\ref{fig:MATanbScan}c) the bounds are relaxed by a factor of about 1.5.

The variation of $\tan\beta$ together with the trilinear soft-breaking parameter $T_u$ is shown in Fig.~\ref{fig:oneloopLQDtanbTu}. The more negative the parameter $(T_u)_{33}$, the larger is the mass splitting between the stop squarks and the lighter is the lightest stop squark. Hence, the limits for the couplings increase (decrease) for decreasing $(T_u)_{33}$ in the case of destructive (constructive) interference. The original benchmark point has $(T_u)_{33}=-3.17~\TeV$ and $\tan\beta=42$. These are the values which we use in all upcoming figures.

We turn now to a discussion of the impact of the different masses appearing in the loop. In general, the case $\lambda^{*\prime}_{ijX}  \lambda_{ij3}^\prime$ ($X=1,2$) is sensitive to different squark and slepton masses. To make this dependence visible we have varied independently the entries $m_{L,ii}^2$ and $m_{Q,jj}^2$ of the soft-breaking masses. The masses that appear in the plots are running $\DRbar$ masses. If a mass is called $m_{\spa t_L}$, this actually means the mass of the mass eigenstate $\spa t_i$ which is mainly $\spa t_L$--like according to the mixing matrix.
\begin{figure}[tpb]
  \centering
  (a) \includegraphics[width=0.45\textwidth]{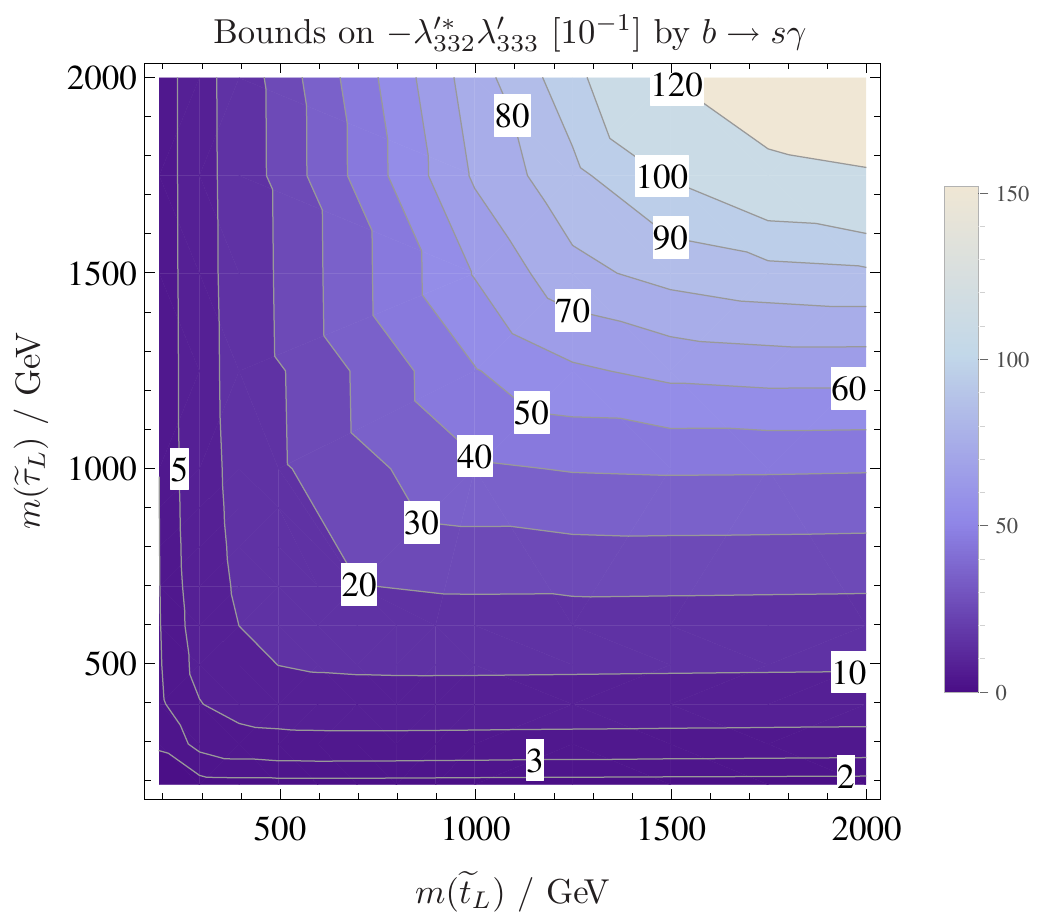} \hfill
  (b) \includegraphics[width=0.45\textwidth]{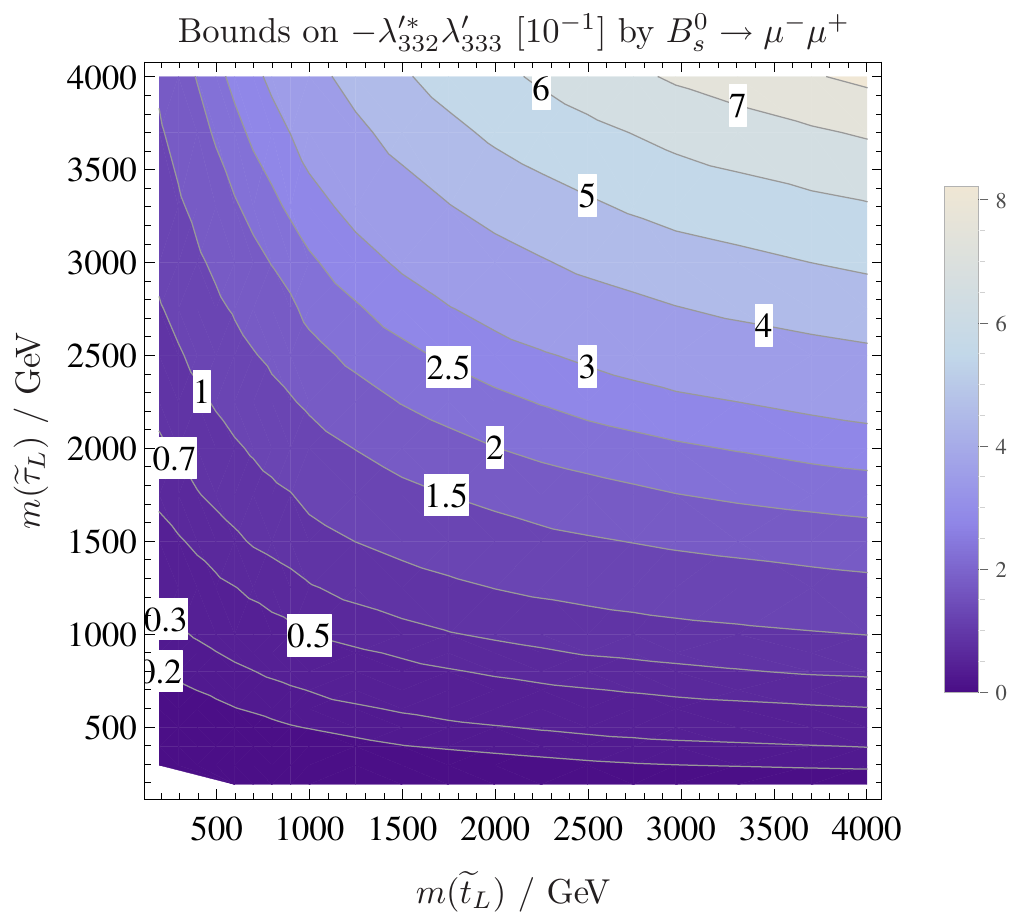} \\
  (c) \includegraphics[width=0.45\textwidth]{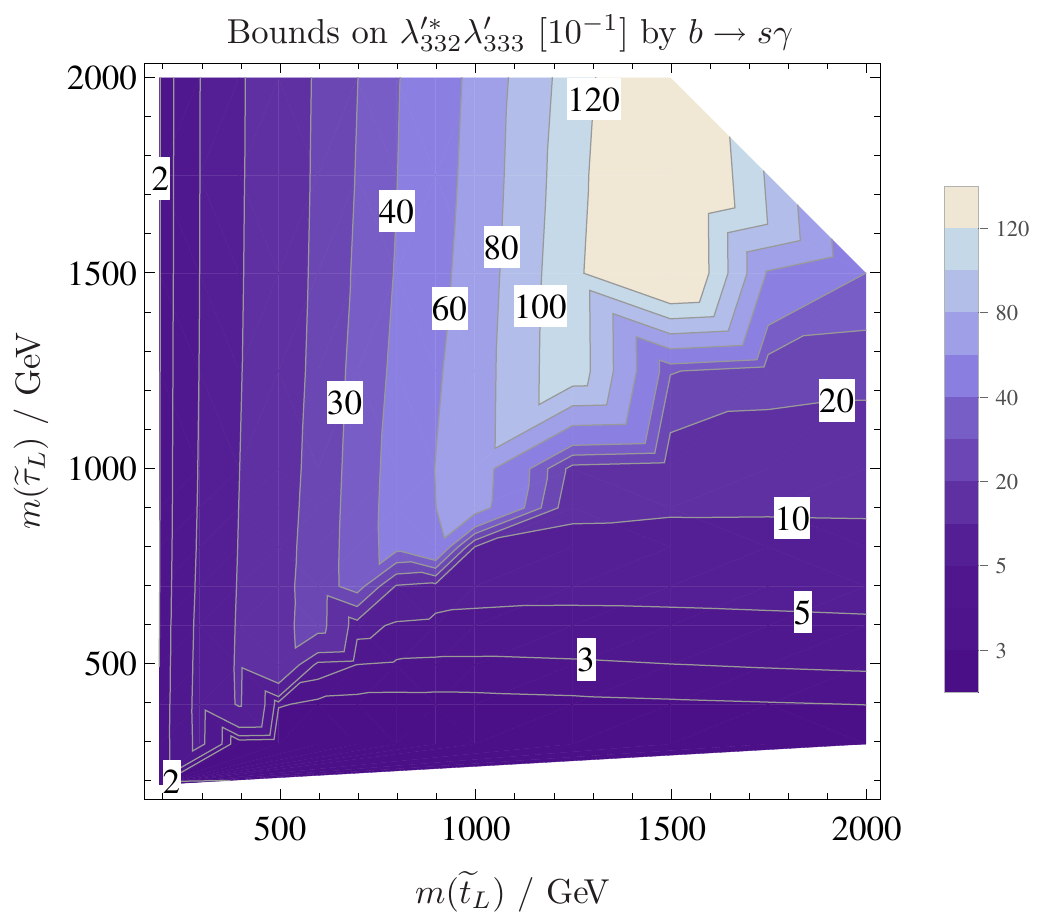} \hfill
  (d) \includegraphics[width=0.45\textwidth]{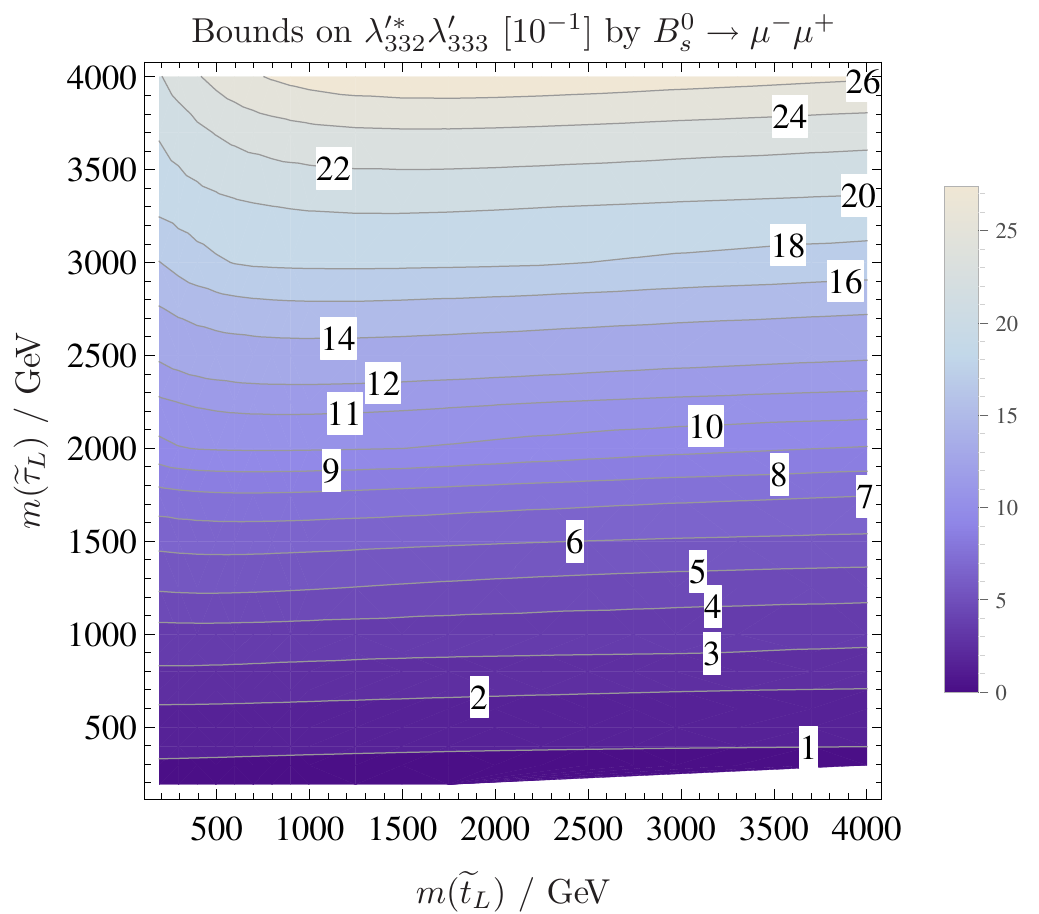}
  \caption{Bounds for $\lpa{332}{333}$ couplings at one-loop with a variation of the masses of $\spa t_L$ and $\spa \tau_L$. First row: Bounds for $\lpa{332}{333}<0$ (constructive interference at $\Bsmm$). Second row: Bounds for $\lpa{332}{333}>0$ (destructive interference for $\Bsmm$). The other parameters are fixed to Tab.~\ref{tab:softparameters}.}
  \label{fig:oneloopLQDijX}
\end{figure}
As an example, we show the upper limit on a specific combination with only third generation sfermions, $\lpa{332}{333}$. In Fig.~\ref{fig:oneloopLQDijX} the upper limits are shown as a function of the masses of the involved top squark ($\spa t_L$) and stau ($\tau_L$). The first row corresponds to a phase $-1$ (constructive interference for $\Bsmm$) and the second row to a phase $+1$ (destructive interference for $\Bsmm$).  The left column shows the bounds from the decay $\bsgamma$, which turn out to be weaker than those from the decay $\Bsmm$ (right column). Interestingly, the choice of sign for constructive interference in $\Bsmm$ is destructive interference for $\bsgamma$ and vice versa. In the case of constructive interference a similar mass dependence can be observed. We can see in Fig.~\ref{fig:oneloopLQDijX}d that there is almost no stop mass dependence, except for heavy stau masses and light stop masses. This can be explained by the scaling behavior in eq.~(\ref{eq:mfscaling}). The $\tau \spa t$ loops become competitive with the $\spa \tau t$ loops only if $m_{\tau} / m_{t} \approx m^2_{\spa t}/m^2_{\spa \tau}$, which applies to the region of the plot where we see a $m_{\spa \tau}$ dependence.

\begin{figure}
  \includegraphics[width=0.49\textwidth]{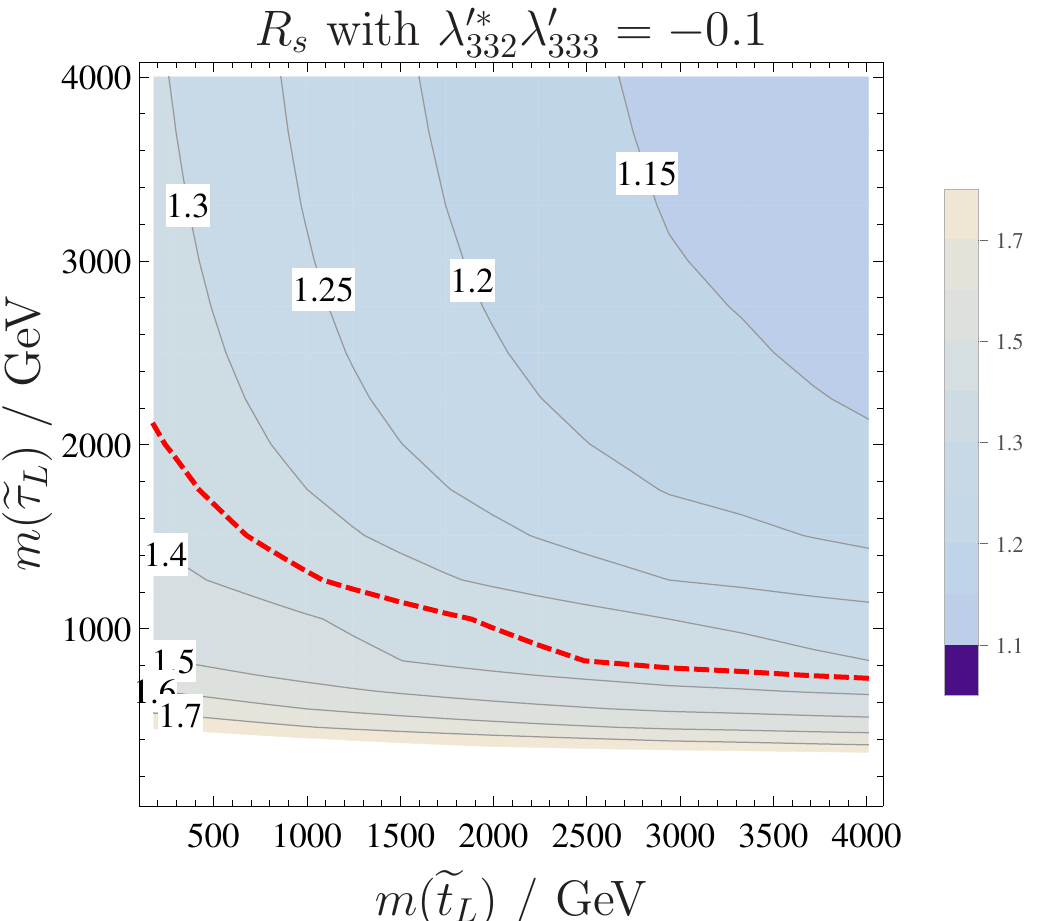} \hfill
\caption{$R_s$ for fixed $\lpa{332}{333}=-0.1$ as function of $m(\tilde{t}_L)$ and $m(\tilde{\tau}_L)$. The area below the red line is excluded by the current experimental limits. }
\label{fig:fixedlambda}
\end{figure}

Finally, we can also check how well $\Bsmm$ has to be measured to constrain the squark masses for a given order of $R$pV couplings. This is done in  Fig.~\ref{fig:fixedlambda} where we plot $R_s$ as function of the involved masses assuming $\lpa{332}{333}=-0.1$. The region below the red dashed line (small squark masses) would then be excluded by the current upper bound on $\BRx{\Bsmm}$. Obviously, if both sfermions are heavier than 2~TeV, the entire contribution is of at most 20\% of the SM contribution. This is the same order of magnitude as the theoretical uncertainty which we have assumed.

\subsubsection{One-Loop Results for {\bf UDD}}
\begin{figure}[tpb]
  \centering
  \includegraphics[width=0.49\textwidth]{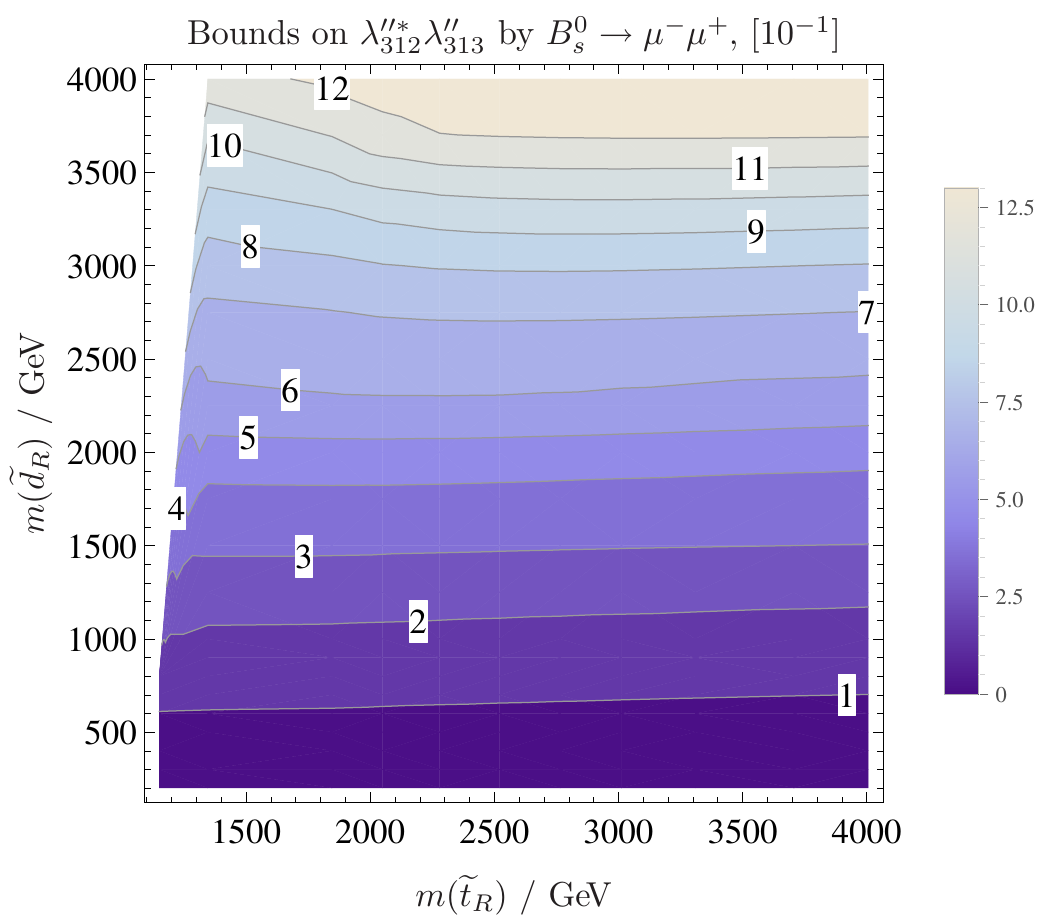} \hfill
  \includegraphics[width=0.49\textwidth]{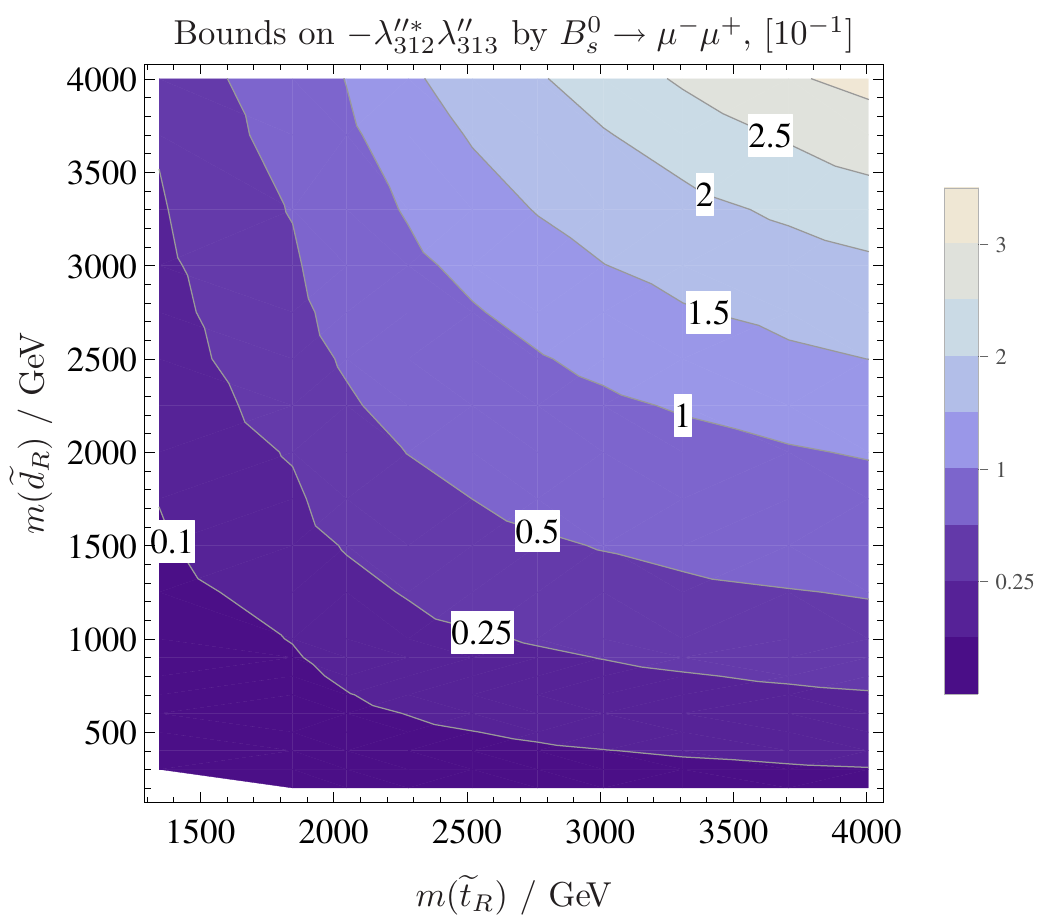} \\
  \caption{Similar to Fig.~\ref{fig:oneloopLQDijX} but for pairs of baryon number violating couplings. On the left we used $\lpp{312}{313}>0$ and on the right $\lpp{312}{313}<0$. }
  \label{fig:oneloopUDD1}
\end{figure}
When considering {\bf UDD} couplings, the masses appearing in the loops from \RpV contributions are the right handed squarks, $\spa d_{Ri}, \spa u_{Ri}$. The stop mass also influences the chargino contribution to $\Bsdmm$, which has already been shown in Fig.~\ref{fig:MATanbScan} (top right) for vanishing \RpV contributions. When \RpV contributions are considered, they come \enquote{on top}. Thus we expect to see slightly better constraints than in the {\bf LQD} case. Indeed, in the case of constructive interference (Fig.~\ref{fig:oneloopUDD1}, right side) the constraints are of the order $10^{-2}$. Like in the {\bf LQD} case, there is almost no down squark mass dependence in Fig.~\ref{fig:oneloopUDD1} (left), except for low stop masses and high $\spa d_{R}$ masses. In contrast, the limits obtained by $\bsgamma$ show nearly the same dependence on $m(\spa d_R)$ and $m(\spa t_R)$ as depicted on the left hand side of Fig.~\ref{fig:oneloopUDD1}. In contrast to {\bf LQD} there is positive interference for $\Bsdmm$ in the case of $\lambda^{\prime\prime*} \lambda^{\prime\prime} < 0$ and destructive interference for positive couplings. $\bsgamma$ shows again exactly the opposite behavior.

%% file: conclusion.tex
\section{Conclusion}
\label{sec:conclusion}
We have presented the first one-loop analysis for $\Bsdmm$ in the MSSM with broken $R$-parity. 
All combinations of couplings between the operators $\lambda \sfd{LLE}$ and $\lambda^\prime\sfd{LQD}$, 
as well as $\lambda^\pprime \sfd{UDD}$ have been considered. We have extracted updated limits 
on combinations which can trigger $B_{s,d}^0 \to \mu \bar{\mu}$ either at tree or one-loop level 
for one benchmark point and discussed the dependence of these limits on the different parameters. 

In this context we have pointed out that a tree level analysis alone might not be sufficient but the bounds and the general behavior can change when including loop-effects due to other SUSY particles. In addition, we presented a set of couplings which lead to `SUSY penguins'. These diagrams rely on an additional flavor change due to the SM or another SUSY loop. Despite this additional suppression the obtained limits can even be stronger than the ones based on direct tree level decays of other mesons because of the strong experimental limits on the $B$-meson sector.
At one-loop level we have shown that only couplings can be constrained by semi-leptonic decays if heavy standard model fermions are involved, i.e. the best limits are stemming from loops involving top quarks. However, even there couplings of $O(1)$ are still in agreement with all observables for SUSY masses in the TeV range and moderate values of $\tan\beta$. $R$pV couplings involving only light SM fermions are not constrained at all by the current measurements of $\Bsdmm$. This is different to $\bsgamma$ where the obtained limits are independent of the generation of the SM particles. Furthermore, we have also shown that the dependence on the phase of the $R$pV couplings is opposite between $\Bsdmm$ and $\bsgamma$, i.e. depending on the sign one has  destructive interference between the SM and $R$pV for one observable and constructive for the other.

%% file: lit.tex
\bibliographystyle{h-physrev}

%% file: boundsUpdatePaper.bbl
\begin{thebibliography}{10}

\bibitem{:2012rz}
{\bf ATLAS Collaboration}, G.~Aad {\em et al.}, ``{Search for
  squarks and gluinos with the ATLAS detector in final states with jets and
  missing transverse momentum using 4.7 $fb^-1$ of sqrt(s) = 7 TeV
  proton-proton collision data},''
  [arXiv:1208.0949 [hep-ex]]

\bibitem{Chatrchyan:2012jx}
{\bf CMS Collaboration}, S.~Chatrchyan {\em et al.}, ``{Search
  for supersymmetry in hadronic final states using MT2 in $pp$ collisions at
  $\sqrt{s} = 7$ TeV},'' 
  {{\em JHEP} {\bf 1210} (2012)  018},
  [arXiv:1207.1798 [hep-ex]]

\bibitem{CMS-PAS-SUS-11-022}
{\bf CMS Collaboration},
``{Search for supersymmetery in final states with missing transverse momentum
  and 0, 1, 2, or $\ge 3$ b jets with CMS},''.

\bibitem{CMS-PAS-SUS-12-005}
{\bf CMS Collaboration},
``{Search for supersymmetry with the razor variables at CMS},''.

\bibitem{:2012mfa}
{\bf CMS Collaboration}, Collaboration, S.~Chatrchyan {\em et al.}, ``{Search
  for new physics in the multijet and missing transverse momentum final state
  in proton-proton collisions at $\sqrt{s} = 7$ TeV},''
 [arXiv:1207.1898 [hep-ex]]

\bibitem{Bechtle:2012zk}
  P.~Bechtle, T.~Bringmann, K.~Desch, H.~Dreiner, M.~Hamer, C.~Hensel, M.~Kr\"amer and N.~Nguyen {\it et al.},
  JHEP {\bf 1206} (2012) 098
  [arXiv:1204.4199 [hep-ph]].

\bibitem{Ghosh:2012dh}
  D.~Ghosh, M.~Guchait, S.~Raychaudhuri and D.~Sengupta,
  Phys.\ Rev.\ D {\bf 86} (2012) 055007
  [arXiv:1205.2283 [hep-ph]].

\bibitem{Buchmueller:2012hv}
  O.~Buchmueller, R.~Cavanaugh, M.~Citron, A.~De Roeck, M.~J.~Dolan, J.~R.~Ellis, H.~Flacher and S.~Heinemeyer {\it et al.},
  Eur.\ Phys.\ J.\ C {\bf 72} (2012) 2243
  [arXiv:1207.7315].


\bibitem{Atlas:2012gk} 
  G.~Aad {\it et al.}  [ATLAS Collaboration],
  Phys.\ Lett.\ B {\bf 716}, 1 (2012)
  [arXiv:1207.7214 [hep-ex]].
  



\bibitem{CMS:2012gu} 
  S.~Chatrchyan {\it et al.}  [CMS Collaboration],
  Phys.\ Lett.\ B {\bf 716}, 30 (2012)
  [arXiv:1207.7235 [hep-ex]].

\bibitem{pMSSM} 
  A.~Arbey, M.~Battaglia, A.~Djouadi and F.~Mahmoudi,
  [arXiv:1211.4004 [hep-ph]].
  M.~W.~Cahill-Rowley, J.~L.~Hewett, A.~Ismail and T.~G.~Rizzo,
  [arXiv:1211.1981 [hep-ph]].
  S.~S.~AbdusSalam,
  [arXiv:1211.0999 [hep-ph]].
    H.~Baer and J.~List,
  [arXiv:1205.6929 [hep-ph]].
    M.~Carena, J.~Lykken, S.~Sekmen, N.~R.~Shah and C.~E.~M.~Wagner,
  Phys.\ Rev.\ D {\bf 86}, 075025 (2012)
  [arXiv:1205.5903 [hep-ph]].

\bibitem{Hall:1983id} 
  L.~J.~Hall and M.~Suzuki,
  Nucl.\ Phys.\ B {\bf 231}, 419 (1984).


\bibitem{rpv}
  B.~C.~Allanach, A.~Dedes and H.~K.~Dreiner,
  Phys.\ Rev.\ D {\bf 69} (2004) 115002
   [Erratum-ibid.\ D {\bf 72} (2005) 079902]
  [hep-ph/0309196].
  H.~K.~Dreiner,
  In *Kane, G.L. (ed.): Perspectives on supersymmetry II* 565-583
  [hep-ph/9707435].
  G.~Bhattacharyya,
  In *Tegernsee 1997, Beyond the desert 1997* 194-201
  [hep-ph/9709395].

\bibitem{Barbier:2004ez} 
  R.~Barbier, C.~Berat, M.~Besancon, M.~Chemtob, A.~Deandrea, E.~Dudas, P.~Fayet and S.~Lavignac {\it et al.},
  Phys.\ Rept.\  {\bf 420}, 1 (2005)
  [hep-ph/0406039].
  
\bibitem{rpvsearches}
  H.~K.~Dreiner and T.~Stefaniak,
  Phys.\ Rev.\ D {\bf 86} (2012) 055010
  [arXiv:1201.5014 [hep-ph]].
  H.~K.~Dreiner and G.~G.~Ross,
  Nucl.\ Phys.\ B {\bf 365} (1991) 597.
M.~Asano, K.~Rolbiecki and K.~Sakurai,
  [arXiv:1209.5778 [hep-ph]].
B.~C.~Allanach and B.~Gripaios,
  JHEP {\bf 1205} (2012) 062
  [arXiv:1202.6616 [hep-ph]].
H.~K.~Dreiner, S.~Grab and T.~Stefaniak,
  Phys.\ Rev.\ D {\bf 84} (2011) 035023
  [arXiv:1102.3189 [hep-ph]].

\bibitem{Franceschini:2012za}
  R.~Franceschini and R.~Torre,
  Eur.\ Phys.\ J.\ C {\bf 73} (2013) 2422
  [arXiv:1212.3622 [hep-ph]].

\bibitem{Evans:2012bf}
  J.~A.~Evans and Y.~Kats,
  JHEP {\bf 1304} (2013) 028
  [arXiv:1209.0764 [hep-ph]].

\bibitem{Dreiner:2012wm}
  H.~K.~Dreiner, F.~Staub, A.~Vicente and W.~Porod,
  Phys.\ Rev.\ D {\bf 86} (2012) 035021
  [arXiv:1205.0557 [hep-ph]].

\bibitem{Barbieri:1993av}
  R.~Barbieri and G.~F.~Giudice,
  Phys.\ Lett.\ B {\bf 309} (1993) 86
  [hep-ph/9303270].

\bibitem{bsgamma1}
  M.~Misiak and M.~Steinhauser,
  Nucl.\ Phys.\ B {\bf 764} (2007) 62
  [hep-ph/0609241];

\bibitem{bsgamma2} 
 M.~Misiak, H.~M.~Asatrian, K.~Bieri, M.~Czakon, A.~Czarnecki, T.~Ewerth, A.~Ferroglia and P.~Gambino {\it et al.},
  Phys.\ Rev.\ Lett.\  {\bf 98} (2007) 022002
  [hep-ph/0609232];
    T.~Hurth, E.~Lunghi and W.~Porod,
  Nucl.\ Phys.\ B {\bf 704} (2005) 56
  [hep-ph/0312260];
    M.~Misiak and M.~Steinhauser,
  Nucl.\ Phys.\ B {\bf 683} (2004) 277
  [hep-ph/0401041];
  C.~Bobeth, P.~Gambino, M.~Gorbahn and U.~Haisch,
  JHEP {\bf 0404} (2004) 071
  [hep-ph/0312090];
H.~K.~Dreiner,
  Mod.\ Phys.\ Lett.\ A {\bf 3} (1988) 867.
\bibitem{BsGamma}
      P.~Gambino and M.~Misiak,
  Nucl.\ Phys.\ B {\bf 611}, 338 (2001)
  [hep-ph/0104034].

\bibitem{bsmumu-susy}
C.~-S.~Huang, W.~Liao and Q.~-S.~Yan,
  Phys.\ Rev.\ D {\bf 59} (1999) 011701
  [hep-ph/9803460].
  K.~S.~Babu and C.~F.~Kolda,
  Phys.\ Rev.\ Lett.\  {\bf 84} (2000) 228
  [hep-ph/9909476].
  C.~Bobeth, T.~Ewerth, F.~Kruger and J.~Urban,
  Phys.\ Rev.\ D {\bf 64} (2001) 074014
  [hep-ph/0104284].
  A.~Dedes, H.~K.~Dreiner and U.~Nierste,
  Phys.\ Rev.\ Lett.\  {\bf 87} (2001) 251804
  [hep-ph/0108037].
  A.~Dedes, J.~Rosiek and P.~Tanedo,
  Phys.\ Rev.\ D {\bf 79} (2009) 055006
  [arXiv:0812.4320 [hep-ph]];
P.~H.~Chankowski and L.~Slawianowska,
  Phys.\ Rev.\ D {\bf 63} (2001) 054012
  [hep-ph/0008046].

\bibitem{deCarlos:1996yh}
  B.~de Carlos and P.~L.~White,
  Phys.\ Rev.\ D {\bf 55} (1997) 4222
  [hep-ph/9609443].
  
\bibitem{Kong:2004cp}
  O.~C.~W.~Kong and R.~D.~Vaidya,
  Phys.\ Rev.\ D {\bf 71} (2005) 055003
  [hep-ph/0403148].

\bibitem{Dreiner:2006gu} 
  H.~K.~Dreiner, M.~Kr\"amer and B.~O'Leary,
  Phys.\ Rev.\ D {\bf 75}, 114016 (2007)
  [hep-ph/0612278].


\bibitem{Li:2013fa}
  C.~Li, C.~-D.~Lu and X.~-D.~Gao,
  [arXiv:1301.3445 [hep-ph]].

\bibitem{Yeghiyan:2013upa}
  G.~Yeghiyan,
  arXiv:1305.0852 [hep-ph].

\bibitem{sarah}
  F.~Staub,
  Computer Physics Communications {\bf 184}, pp. 1792 (2013)
  [arXiv:1207.0906 [hep-ph]].
  F.~Staub,
  Comput.\ Phys.\ Commun.\  {\bf 182} (2011) 808
  [arXiv:1002.0840 [hep-ph]].
  F.~Staub,
  Comput.\ Phys.\ Commun.\  {\bf 181} (2010) 1077
  [arXiv:0909.2863 [hep-ph]].
  F.~Staub,
  [arXiv:0806.0538 [hep-ph]].

\bibitem{spheno}
  W.~Porod and F.~Staub,
  Comput.\ Phys.\ Commun.\  {\bf 183} (2012) 2458
  [arXiv:1104.1573 [hep-ph]].
  W.~Porod,
  Comput.\ Phys.\ Commun.\  {\bf 153} (2003) 275
  [hep-ph/0301101].
  
\bibitem{Dreiner:2012dh}
  H.~Dreiner, K.~Nickel, W.~Porod and F.~Staub,
  [arXiv:1212.5074 [hep-ph]].

\bibitem{Dedes:2008iw}
  A.~Dedes, J.~Rosiek and P.~Tanedo,
  Phys.\ Rev.\ D {\bf 79} (2009) 055006
  [arXiv:0812.4320 [hep-ph]].


\bibitem{Buras:2012ru} 
  A.~J.~Buras, J.~Girrbach, D.~Guadagnoli and G.~Isidori,
  [arXiv:1208.0934 [hep-ph]].

\bibitem{Raven:2012fb}
  G.~Raven [LHCb Collaboration],
  arXiv:1212.4140 [hep-ex].

\bibitem{Buras:2013uqa}
  A.~J.~Buras, R.~Fleischer, J.~Girrbach and R.~Knegjens,
  [arXiv:1303.3820 [hep-ph]].



\bibitem{LHCb:2012ct} 
  LHCb Collaboration: R.~Aaij and many more,
  Phys.\ Rev.\ Lett.\  {\bf 110} (2013) 021801
  [arXiv:1211.2674 [hep-ex]].


\bibitem{LHCb:2013}
  LHCb Collaboration: R.~Aaij, B.~Adeva, and many more,
  arXiv:1307.5024

  \bibitem{Haisch:2012re} 
  U.~Haisch and F.~Mahmoudi,
  JHEP {\bf 1301}, 061 (2013)
  [arXiv:1210.7806 [hep-ph]].



\bibitem{Beringer:1900zz} 
  J.~Beringer {\it et al.}  [Particle Data Group Collaboration],
  Phys.\ Rev.\ D {\bf 86}, 010001 (2012).

\bibitem{Dreiner:2003hw} 
  H.~K.~Dreiner and M.~Thormeier,
  Phys.\ Rev.\ D {\bf 69}, 053002 (2004)
  [hep-ph/0305270].

\bibitem{Dreiner:2012mx} 
  H.~K.~Dreiner, K.~Nickel, F.~Staub and A.~Vicente,
  Phys.\ Rev.\ D {\bf 86}, 015003 (2012)
  [arXiv:1204.5925 [hep-ph]].







\bibitem{Lunghi:2006hc}
  E.~Lunghi and J.~Matias,
  JHEP {\bf 0704} (2007) 058
  [hep-ph/0612166].

\bibitem{Staub:2011dp} 
  F.~Staub, T.~Ohl, W.~Porod and C.~Speckner,
  Comput.\ Phys.\ Commun.\  {\bf 183}, 2165 (2012)
  [arXiv:1109.5147 [hep-ph]].
  

\bibitem{Cahill-Rowley:2013gca}
  M.~W.~Cahill-Rowley, J.~L.~Hewett, A.~Ismail, M.~E.~Peskin and T.~G.~Rizzo,
  [arXiv:1305.2419 [hep-ph]].

 
\bibitem{Berger:2008cq} 
  C.~F.~Berger, J.~S.~Gainer, J.~L.~Hewett and T.~G.~Rizzo,
  JHEP {\bf 0902}, 023 (2009)
  [arXiv:0812.0980 [hep-ph]].
  
\bibitem{Camargo-Molina:2013qva} 
  J.~E.~Camargo-Molina, B.~O'Leary, W.~Porod and F.~Staub,
  arXiv:1307.1477 [hep-ph].
  
\bibitem{Buras:2002vd} 
  A.~J.~Buras, P.~H.~Chankowski, J.~Rosiek and L.~Slawianowska,
  Nucl.\ Phys.\ B {\bf 659}, 3 (2003)
  [hep-ph/0210145].
\end{thebibliography}
